\newcommand{\rone}{FRB~20121102A\xspace}

\newcommand{\evnburstslband}{$\alpha=05^{\mathrm{h}}31^{\mathrm{m}}58.7016^{\mathrm{s}}$, $\delta = +33\degr08\arcmin52.5483\arcsec$ (J2000, ICRF)}

\newcommand{\evnburstslbanderror}{
\begin{align*}
    \alpha_{\mathrm{FRB}} &= 05^{\mathrm{h}}31^{\mathrm{m}}58.7016^{\mathrm{s}} \pm 2.5\,\mathrm{mas} \, \mathrm{(J2000, ICRF)}\\
    \delta_{\mathrm{FRB}} &= +33\degr08\arcmin52.5483\arcsec \pm 2.5\,\mathrm{mas} \, \mathrm{(J2000, ICRF)}
\end{align*}
}

\newcommand{\evnprslbanderror}{
\begin{align*}
    \alpha_{\mathrm{PRS}} &= 05^{\mathrm{h}}31^{\mathrm{m}}58.7015^{\mathrm{s}} \pm 2.5\,\mathrm{mas} \, \mathrm{(J2000, ICRF)}\\
    \delta_{\mathrm{PRS}} &= +33\degr08\arcmin52.5518\arcsec \pm 2.5\,\mathrm{mas} \, \mathrm{(J2000, ICRF)}
\end{align*}
}

\newcommand{\evnprsCband}{$\alpha=05^{\mathrm{h}}31^{\mathrm{m}}58.70159^{\mathrm{s}}$, $\delta = +33\degr08\arcmin52.550100\arcsec$ (J2000, ICRF)}

\newcommand{\evncorrpos}{$\alpha=05^{\mathrm{h}}31^{\mathrm{m}}58.7^{\mathrm{s}}$, $\delta = +33\degr08\arcmin52.57\arcsec$ (J2000, ICRF)}

\newcommand{\MU}{Department of Physics, McGill University, 3600 rue University, Montr\'eal, QC H3A 2T8, Canada}
\newcommand{\TSI}{Trottier Space Institute, McGill University, 3550 rue University, Montr\'eal, QC H3A 2A7, Canada}

\newcommand{\LPCE}{LPC2E, OSUC, Univ Orleans, CNRS, CNES, Observatoire de Paris, F-45071 Orleans, France}
\newcommand{\ORN}{ORN, Observatoire de Paris, Universit\'e PSL, Univ Orl\'eans, CNRS, 18330 Nan\c{c}ay, France}
\newcommand{\MPIfR}{Max-Planck-Institut für Radioastronomie, Auf dem Hügel 69, 53121 Bonn, Germany}
\newcommand{\INAFBologna}{INAF-Istituto di Radioastronomia, Via Gobetti 101, 40129, Bologna, Italy}
\newcommand{\INAFCagliari}{INAF-Osservatorio Astronomico di Cagliari, via della Scienza 5, I-09047, Selargius (CA), Italy}
\newcommand{\INAFCatania}{INAF-Osservatorio Astrofisico di Catania, via Santa Sofia 78, I-95123, Catania, Italy}
\newcommand{\VUAS}{Engineering Research Institute Ventspils International Radio Astronomy Centre (ERI VIRAC) of Ventspils University of Applied Sciences, Inzenieru street 101, Ventspils, LV-3601, Latvia}

\newcommand{\IAUMK}{Institute of Astronomy, Faculty of Physics, Astronomy and Informatics, Nicolaus Copernicus University, Grudziadzka 5, PL-87-100 Toru\'n, Poland}

\newcommand{\Chalmers}{Department of Space, Earth and Environment, Chalmers University of Technology, Onsala Space Observatory, 439 92, Onsala, Sweden}
\newcommand{\API}{Anton Pannekoek Institute for Astronomy, University of Amsterdam, Science Park 904, 1098 XH, Amsterdam, The Netherlands}
\newcommand{\ASTRON}{ASTRON, Netherlands Institute for Radio Astronomy, Oude Hoogeveensedijk 4, 7991 PD Dwingeloo, The Netherlands}
\newcommand{\JIVE}{Joint Institute for VLBI ERIC, Oude Hoogeveensedijk 4, 7991~PD Dwingeloo, The Netherlands}
\newcommand{\Bristol}{School of Physics, University of Bristol, Tyndall Avenue, Bristol BS8 1TL, UK}
\newcommand{\JBCA}{Jodrell Bank Centre for Astrophysics, Dept. of Physics \& Astronomy, University of Manchester, Manchester M13 9PL, UK}

\newcommand{\MITK}{MIT Kavli Institute for Astrophysics and Space Research, Massachusetts Institute of Technology, 77 Massachusetts Ave, Cambridge, MA 02139, USA}

\newcommand{\SKAO}{SKA Observatory (SKAO), Science Operations Centre, CSIRO ARRC, Kensington WA 6151, Australia}
\newcommand{\STANFORD}{Department of Physics, Stanford University, 382 Via Pueblo Mall, Stanford, CA 94305, USA}
\newcommand{\KIPAC}{Kavli Institute for Particle Astrophysics and Cosmology, Stanford University, 382 Via Pueblo Mall, Stanford, CA 94305, USA}

\documentclass[twocolumn]{aa}

\usepackage{xcolor}
\usepackage{graphicx}
\usepackage{txfonts}

\usepackage[
    colorlinks=true,
    linkcolor=blue,
    citecolor=blue,
    filecolor=blue,
    urlcolor=blue
]{hyperref}

\usepackage{upgreek}

\usepackage{siunitx}
\usepackage{booktabs}

\begin{document}

\makeatletter
\let\linenumbers\relax
\let\pagewiselinenumbers\relax
\let\runninglinenumbers\relax
\let\switchlinenumbers\relax
\let\resetlinenumber\relax
\let\setrunninglinenumbers\relax
\let\modulolinenumbers\relax
\let\linenumberfont\relax
\makeatother

\title{Revisiting \rone: milliarcsecond localisation and a decreasing dispersion measure}
\titlerunning{Revisiting \rone}

\subtitle{} 

\author{
M.P.~Snelders \inst{1,2}\thanks{{Email: snelders@astron.nl \\ \href{https://orcid.org/0000-0001-6170-2282}{ORCID: 0000-0001-6170-2282}}}
\and
J.W.T.~Hessels \inst{3,4,2,1}
\and
J.~Huang \inst{3,4}
\and
N.~Sridhar \inst{5,6}
\and
B.~Marcote \inst{7,1}
\and
A.M.~Moroianu \inst{2}
\and
O.S.~Ould-Boukattine \inst{1,2}
\and
F.~Kirsten \inst{8,1}
\and
S.~Bhandari \inst{9}
\and
D.M.~Hewitt \inst{2}
\and
D.~Pelliciari \inst{10}
\and
L.~Rhodes \inst{3,4}
\and
R.~Anna-Thomas \inst{1,2}
\and \\
U.~Bach \inst{11}
\and
E.K.~Bempong-Manful \inst{12,13}
\and
V.~Bezrukovs \inst{14}
\and
J.D.~Bray \inst{12}
\and
S.~Buttaccio \inst{15,16}
\and
I.~Cognard \inst{17,18}
\and \\
A.~Corongiu \inst{19}
\and
R.~Feiler \inst{20}
\and
M.P.~Gawro\'nski \inst{20}
\and
M.~Giroletti \inst{10}
\and
L.~Guillemot \inst{17,18}
\and
R.~Karuppusamy \inst{11}
\and \\
M.~Lindqvist \inst{8}
\and
K.~Nimmo \inst{21}
\and
A.~Possenti \inst{19}
\and
W.~Puchalska \inst{20}
\and
D.~Williams-Baldwin \inst{12}
}

\institute{
\ASTRON
\and
\API
\and
\MU
\and
\TSI
\and
\STANFORD
\and
\KIPAC
\and
\JIVE
\and
\Chalmers
\and
\SKAO
\and
\INAFBologna
\and
\MPIfR
\and
\JBCA
\and
\Bristol
\and
\VUAS
\and
\INAFCatania
\and
\INAFBologna
\and
\LPCE
\and
\ORN
\and
\INAFCagliari
\and
\IAUMK
\and
\MITK
}

\date{Received Month XX, 202X; accepted Month XX, 202X}

 
\abstract{
\rone is the original repeating fast radio burst (FRB) source and also the first to be localised to milliarcsecond precision using very-long-baseline interferometry (VLBI). It has been active for over 13 years and resides in an extreme magneto-ionic environment in a dwarf host galaxy at a distance of $\sim$$1$\,Gpc. In this work, we use the European VLBI Network (EVN) to (re-)localise \rone and its associated persistent radio source (PRS). We confirm that the two are co-located --- improving on previous results by a factor of $\sim$$4$ and constraining the FRB and PRS co-location to $\sim$$12$\,pc transverse offset. Over a decade, the PRS luminosity on milliarcsecond scales remains consistent with measurements on larger angular scales, showing that the PRS is still compact. We also present the detection of $18$ bursts with the Nan\c{c}ay Radio Telescope (NRT) as part of our \'ECLAT monitoring program. These bursts, together with previously published results, show that the observed dispersion measure (DM) of \rone has dropped by $\sim$$25$\,pc\,cm$^{-3}$ in the past five years, highlighting a fractional decrease in the local DM contribution of $\gtrsim$$15$\,\%. We discuss potential physical scenarios and highlight possible future observations that will help reveal the nature of \rone, which is one of only a few known FRBs with a luminous PRS.
}


\maketitle

\clearpage

\section{Introduction}\label{sec:intro}

Fast radio bursts (FRBs) are millisecond-duration, extragalactic flashes of coherent radio waves whose origin(s) and emission mechanism(s) are still poorly understood \citep[see, e.g.,][for a review]{petroff_2022_aapr}. A small fraction ($\sim$$3$\,\%; e.g., \citealt{spitler_2016_nature,chime_2023_apjl}) of the known FRB population ($\gtrsim$$3,000$ sources; e.g., \citealt{wang_2025_arxiv_cat2}) has been observed to repeat, implying that these sources have longer-lived central engines that are capable of producing repeat bursts on timescales from minutes to several years. It remains unclear if all FRBs repeat and have the same progenitors (see \citealt{pleunis_2021_apj} and also, e.g., \citealt{ouldboukattine_2024_arxiv,chime_2025_apjl_rbfloat}, and references therein).

In this work, we focus on \rone, the first known repeater \citep{spitler_2016_nature}, and one of the best-studied FRB sources. \rone was precisely localised, using very-long-baseline interferometry \citep[VLBI;][]{chatterjee_2017_nature,Marcote_2017_ApJL}, to a low-metallicity, star-forming dwarf galaxy at a redshift of $z \sim 0.193$ \citep{tendulkar_2017_apjl,bassa_2017_apjl}. 
At a position that is consistent with the bursts, a persistent radio source (PRS) was found \citep{chatterjee_2017_nature}; this source remains compact in inter-continental VLBI observations with the European VLBI Network (EVN) and has a luminosity of $\sim$$5\times10^{38}$\,erg\,s$^{-1}$ \citep[][]{Marcote_2017_ApJL}. The transverse size of the PRS is $\lesssim$$0.7$\,pc (diameter) and the projected separation between the burst source and the PRS is $\lesssim$$40$\,pc \citep{Marcote_2017_ApJL}. \rone's co-location with this luminous PRS makes it a peculiar source compared to the vast majority of the FRB population, with the closest-known analogue being FRB~20190520B \citep{niu_2022_nature,bhandari_2023_apjl}. Whether these sources represent an evolutionary stage common to repeaters in general, or instead a distinct subclass, is unclear. The PRS provides important insight into the nature of \rone and its local environment --- as we discuss in detail in Section~\ref{sec:discussion}, where we consider four possible PRS origins: 1) a nebula powered by a magnetar; 2) an accreting binary; 3) a magnetically interacting binary; and 4) a `wandering' massive black hole.

\rone is unreliably detectable, making follow-up and monitoring observations challenging. Its burst rate changes on timescales of hours to years and its activity may be periodic at $159.3$\,days --- albeit with a $>$$50$\,\% duty cycle \citep[see, e.g.,][and references therein]{braga_2025_arxiv}. 

Nonetheless, monitoring of propagation effects incurred by the bursts also provides valuable insights. For example, both the dispersion measure (DM) and Faraday rotation measure (RM) of the \rone bursts have varied wildly in the past decade. The DM first increased by $\sim$$6$\,pc\,cm$^{-3}$ before dropping by $\sim$$20$\,pc\,cm$^{-3}$ (\citealt{wang_2025_arxiv}). These changes must be in the local environment, where DM$_{\textrm{host$+$local}}$ is estimated to be $< 225$\,pc\,cm$^{-3}$ \citep{tendulkar_2017_apjl} and depends on how much DM should be ascribed to the interstellar medium (ISM) of the host. Thus, the fractional change in DM$_{\textrm{host$+$local}}$ must be large, $\gtrsim10$\,\%. This behaviour is not clearly seen in other repeating FRBs. 

\rone also shows an exceptionally large and highly variable RM\footnote{Each RM value is quoted in the corresponding source rest frame. To convert it to the observer frame simply divide the value by $(1+z)^{2}$, where $z$ is the redshift of the host galaxy of \rone.}, which was $\sim$$181 \times 10^{3}$ \,rad\,m$^{-2}$ in 2016, and by 2023 dropped to $\sim$$44 \times 10^{3}$ \,rad\,m$^{-2}$ \citep[][]{wang_2025_arxiv,Michilli_2018_Nature,Plavin_2022_MNRAS,hilmarsson_2021_apjl}. Even after the RM decreased by $\sim$$75$\,\%, it is still higher than the next-highest |RM| FRB \citep[FRB~20190520B;][]{annathomas_2023_science} and several orders of magnitude larger than typical FRB RMs \citep[e.g.,][]{pandhi_2024_apj}. The high RM, which is likely caused by the local environment of the source, combined with its variability and the apparently associated PRS, together indicate that \rone resides in an extreme and dynamic magneto-ionic environment \citep{Michilli_2018_Nature}. 

Associations between (repeating) FRB sources and PRSs remain rare. Besides \rone, there is the previously mentioned FRB~20190520B. Like \rone, this source is also localised to a dwarf galaxy and shows large RM variations, including sign flips \citep{niu_2022_nature,annathomas_2023_science,bhandari_2023_apjl}. Recently, it was confirmed that FRB~20190417A is also associated with a PRS \citep{ibik_2024_apj,moroianu_2025_arxiv}. The host of this source is also a dwarf galaxy and the bursts show large RM variations of $\sim$$10^{3}$ \,rad\,m$^{-2}$\,month$^{-1}$ ($\sim$$20$\,\% fractional change). Other proposed FRB-PRS systems associated with FRB~20201124A and FRB~20240114A are much less luminous and with much lower RMs \citep{bruni_2024_nature,bruni_2025_aap}.

Continued monitoring of \rone and its nearby PRS are key to deciphering the physical nature of this system. While much has been gleaned from the observations to date, there are important missing pieces. \citet{Marcote_2017_ApJL} showed that the associated PRS is compact on sub-milliarcsecond scales; however, due to limited $(u,v)$-coverage in their VLBI observations they were only able to constrain the physical offset between the burst source and PRS to $<$$40$\,pc --- leaving open the possibility that the two sources might be slightly offset from each other. Likewise, while several telescopes have monitored the flux density of the PRS \citep[e.g.,][]{rhodes_2023_mnras,bhardwaj_2025_arxiv}, there are few VLBI observations to verify whether the flux density tracks in the same way on milliarcsecond scales, and whether a substantial fraction of the flux density resolves out on long baselines.

Motivated to robustly establish as much as possible about the PRS, we observed \rone with the EVN, triggered by a report of enhanced burst activity \citep{wang_2022_atel}. In this paper, we present the results of our EVN campaign towards \rone, and place a strong constraint on the co-location of the burst source and PRS. We also present the results of our \rone monitoring campaign with the Nan\c{c}ay Radio Telescope (NRT). We discuss possible scenarios that explain the FRB and PRS association, combined with the evolving DM and RM of the source.

\section{Observations}\label{sec:obs}
\subsection{EVN observations}\label{subsec:evn_obs}
We observed \rone at four epochs in 2022/2023 (Table~\ref{tab:obs_log}) using an \textit{ad-hoc} sub-array of EVN dishes, as part of the PRECISE project (Pinpointing REpeating ChIme Sources with EVN dishes; PI: F.~Kirsten). All observing runs were carried out between $1254$--$1510$\,MHz in a mixed-frequency setup, where every station records either two, four, six or eight 32-MHz subbands. All stations recorded coherent channelised voltages as $2$-bit samples, storing both left and right circular polarisations in VDIF format \citep{Whitney_2010_ivs}.

We used phase referencing with a cycle time of $\sim$$8$\,minutes, where $\sim$$6$-minute scans on \rone are interleaved with $\sim$$2$-minute phase calibrator scans on J0529+3209 ($1.1\degr$ away from \rone, hereafter called the `phase calibrator'). Additionally, we observed J0541+3301 (hereafter called `the check source') roughly twice per hour, $3$\,minutes per scan, and used it as a check source to verify the astrometric solutions. The check source is $2.1\degr$ away from \rone and $2.7\degr$ away from the phase calibrator. The source J1829+4844 was used as a fringe finder and bandpass calibrator. The pulsars B0329+54, B0540+23, and B1933+16 were observed to verify the data quality, frequency setup, polarimetric calibration, and burst-search pipeline. The observation and calibration strategy of the first two epochs are illustrated in Appendix~\ref{app:obs_timeline}.

\begin{table}[ht]
    \centering
    \caption{Observations log.}
    \label{tab:obs_log}
    \begin{tabular}{lccc}
        \hline
        PRECISE/EVN              & Observation               & Num. of  & Duration                    \\
        codes$^{\text{a}}$       & MJD                      & bursts   & [min.]$^{\text{b}}$ \\
        \hline
        PR242A/EK051E           & $59843$--$59844$           & 2        & 210           \\
        PR243A/EK051F           & $59849$--$59850$           & 5        & 276           \\
        PR268A\phantom{/EK051X} & $60022$\phantom{--$60022$}  & 1        & \phantom{0}68 \\
        PR269A\phantom{/EK051X} & $60028$--$60029$ & 0        & 138           \\
        \hline
        \multicolumn{4}{l}{$^{\text{a}}$ The data of observations PR268A and PR269A were} \\
        \multicolumn{4}{l}{\quad not correlated and thus do not have an EVN} \\
        \multicolumn{4}{l}{\quad experiment code (Section~\ref{subsec:interferometric_data}).} \\
        \multicolumn{4}{l}{$^{\text{b}}$ Defined as the on-source time with the Effelsberg telescope.} \\
    \end{tabular}
\end{table}

\subsection{NRT observations}\label{subsec:nrt_obs}
\rone is regularly monitored as part of the \'ECLAT (Extragalactic Coherent Light from Astrophysical Transients; PI: D.~Hewitt) observing campaign on the NRT (in France). Between February 6, 2022 and June 6, 2025, \rone has been observed $125$ times for a total of $106$\,hours (see Appendix~\ref{app-sec:nrt_bursts_logs}). Most observations take place at a central frequency of $1484$\,MHz (L-band) with $512$\,MHz of bandwidth, with the occasional observation centred at $\sim$$2.5$\,GHz (S-band; Appendix~\ref{app-sec:nrt_bursts_logs}). The Nan\c{c}ay Ultimate Pulsar Processing Instrument (NUPPI; \citealt{Desvignes_2011}) records full-polarisation data (in a linear basis) with $16$-$\upmu$s time resolution and $4$-MHz channels, and writes out the samples as $32$-bit floating point numbers. During the observation, coherent dedispersion
(i.e., dedispersion within spectral channels) is applied, where we use an MJD-dependent coherent dispersion measure, DM$_{\mathrm{NRT,coherent}}$ (Appendix~\ref{app-sec:nrt_bursts_logs}).

\section{Analysis}\label{sec:analysis}
\subsection{Search for bursts}
\subsubsection{Effelsberg radio telescope data}

After every PRECISE/EVN observation, we transferred the voltage data from the Effelsberg radio telescope to a dedicated server at the Onsala Space Observatory (Sweden). The voltages were converted to Stokes~I (intensity) filterbank \citep{Lorimer_2011_ascl} data with a time and frequency resolution of $64$\,$\upmu$s and $31.25$\,kHz, respectively, using \texttt{digifil} \citep{vanStraten_2011_PASA}. We searched the data for FRBs using \texttt{Heimdall} \citep{Heimdall} with a DM range of $557 \pm 50$\,pc\,cm$^{-3}$. Single-pulse candidates found by \texttt{Heimdall} were further analysed with \texttt{FETCH} \citep{Agarwal_2020_MNRAS}, which is a deep-learning-based classifier that aims to distinguish astrophysical signals from radio frequency interference (RFI). All candidates for which \texttt{FETCH} assigns a $\geq0.5$ probability that the signal is astrophysical are manually inspected. The search pipeline is described in more detail in \citet{kirsten_2021_natas,kirsten_2022_natur}. We detected two bursts (A$1$, A$2$) at Epoch~1, five bursts (B$1$--B$5$) at Epoch~2, one burst (C$1$) at Epoch~3 and no bursts at Epoch~4 (Figure~\ref{fig:family_plot} and Table~\ref{tab:obs_log}).

\subsubsection{NRT data}
The full-polarisation NUPPI data is converted to $8$-bit Stokes~I filterbank data with \texttt{digifil}, without downsampling in time or frequency. These data are passed, as $2$-minute chunks, to \texttt{rfifind} from \texttt{PRESTO} \citep{presto} to generate an RFI mask that flags badly affected frequency channels. Temporal masking is not applied. We then used \texttt{Heimdall} to search these filterbank files, with a DM range of $\pm 25$\,pc\,cm$^{-3}$ around the DM$_{\mathrm{NRT,coherent}}$ (Appendix~\ref{app-sec:nrt_bursts_logs}) value of that observation, for single-pulses that exceed a S/N of $7$. Similar to the Effelsberg radio telescope data, all candidates where \texttt{FETCH} assigned a $\geq0.5$ probability of being astrophysical were manually inspected. In total, we found $18$ bursts, which are labelled N$01$--N$18$ in Appendix~\ref{app-sec:nrt_bursts_logs}. Snippets around the burst times were extracted from the $32$-bit full-polarimetric data. The NRT burst search pipeline is fully described in \citet{hewitt_2023_mnras}.

\subsection{Interferometric data}
\label{subsec:interferometric_data}
We chose not to correlate the data from the PRECISE/EVN observation PR268A because only one weak burst was found, the observation was short (Table~\ref{tab:obs_log}), and it was strongly affected by RFI. Observation PR269A was not correlated because no bursts were found. The data of all participating stations from observations PR242A (Epoch~1) and PR243A (Epoch~2) were transferred to the Joint Institute for VLBI ERIC (JIVE, in the Netherlands) for correlation with the FX Software Correlator \texttt{SFXC} \citep{Keimpema_2015_ExA}. The participating stations were: Effelsberg (Ef; Germany), Toru\'n (Tr; Poland), Onsala (O8; Sweden), Westerbork RT-1 (Wb; Netherlands), Noto (Nt; Italy), Medicina (Mc; Italy, only participated at Epoch~1), Irbene (Ir; Latvia) and six e-MERLIN stations (Cambridge (Cm), Darnhall (Da), Defford (De), Knockin (Kn), Pickmere (Pi), and Jodrell Bank Mark II (Jm); United Kingdom). The calibrator data and continuum data of \rone were correlated using $2$-s integrations and $0.5$-MHz channel widths ($64$ channels per $32$-MHz subband). The continuum data of \rone, as well as the burst data, were referenced to a phase centre of \evncorrpos, which is $\sim$$30$\,mas (milliarcseconds) away from the previously published C-band PRS position \citep[\evnprsCband][]{Marcote_2017_ApJL}. The correlation phase centre and the C-band PRS position are shown in Figure~\ref{fig:bursts_dirty_zoom}. The burst data (A$1$, A$2$ and B$1$--B$5$) were coherently dedispersed and correlated using various gates that depend on the temporal widths and arrival times of the bursts and are also channelized to $64$ channels per subband. All output from the \texttt{SFXC} correlator is saved as \texttt{FITS-IDI} (Flexible Image Transport System Interferometry Data Interchange) files \citep[e.g.,][]{AIPS}.

We processed and calibrated the correlated continuum data of the calibrators and the PRS associated with \rone using \texttt{AIPS} \citep{AIPS}. Standard EVN calibration tables were used, which contain the parallactic angle correction and \textit{a-priori} gain correction, using the gain curves and system temperature measurements that the stations recorded during the observations. We also applied the EVN flagging table, ignoring all data from antennas that were still slewing. Additionally, we flagged $\sim$$15$\,\% ($5$ channels on each side) at the edges of the subbands, where antenna sensitivity significantly drops. The \texttt{AIPS} task \texttt{VLBATECR} was used to correct for the ionospheric dispersive delays \citep{Petrov_2023_AJ}. We note however that, in this case, omitting the \texttt{VLBATECR} step results in no measurable shift in the \rone PRS position but increases the flux density (Section~\ref{sec:results}) by $\sim$$3$\,\% (well within the measured uncertainties).

We used the \texttt{AIPS} tool \texttt{spflg} to manually flag the visibilities of the fringe finder, phase calibrator, check source and the continuum data of \rone as a function of time, frequency, and baseline. At the start of the calibration the scan on the bright fringe finder was used to correct for the instrumental delay between subbands (i.e., using \texttt{fring} without fitting for a rate), with the assumption that that delay remains constant during the observation. Next, the \texttt{AIPS} task \texttt{fring} was used on all scans of the phase calibrator to determine the group delay and phase rate (i.e., the phase solutions) as a function of time and frequency. Finally, the \texttt{AIPS} task \texttt{bpass} was used on the fringe finder scan. The calibrated and flagged visibilities of both epochs were concatenated and imaged in \texttt{Difmap} \citep{difmap} to create a model of the phase calibrator that was used for self calibration. The calibration solutions were transferred and interpolated to the \rone continuum data. The flagged, concatenated, and calibrated visibilities were exported as \texttt{UVFITS} files with \texttt{fittp}, imported in \texttt{CASA} with \texttt{importuvfits} and imaged with \texttt{tclean}. 

The burst visibilities were calibrated differently. We started by importing all \texttt{FITS-IDI} files of the bursts and calibrator data in \texttt{CASA} using \texttt{importfitsidi}. Next, we used \texttt{split} to create a \texttt{CASA} measurement set that only contains the visibilities of the scans of the fringe finder, the bursts, and the $13$ phase calibrator scans that bracket the bursts. This results in a much smaller dataset that allows for quicker processing and manual RFI flagging. Similar to the continuum data that was processed with \texttt{AIPS}, we flagged $\sim$$15$\,\% of the subband edges and we corrected for the ionosphere with \texttt{CASA}'s \texttt{gencal} and \texttt{tec\_maps} (i.e., the \texttt{AIPS} equivalent of \texttt{VLBATECR}). We manually flagged all time ranges for which a particular antenna was still slewing and flagged for RFI in the fringe finder scans as a function of time, frequency and baseline. We did the same for each phase calibrator bracket, keeping track of which frequency channels, as a function of baseline, were flagged. For the individual bursts in such a bracket, we flagged the same frequency channels. This approach was chosen because each burst has only one time integration, making it difficult to distinguish between strong scintillation, narrowband RFI and instrumental effects. We used \texttt{CASA}'s \texttt{fringefit}, without fitting for fringe rates, on the fringe finder scans to correct for the instrumental delay between subbands and on the phase calibrator scans to determine the phase, delay, and delay rate as a function of time and frequency. Finally, \texttt{CASA}'s \texttt{bandpass} was used to determine the antenna response, for both phase and amplitude, as a function of frequency. The calibration solutions were transferred to the flagged burst visibilities. Similar to the continuum data, we made use of \texttt{tclean} to create dirty maps (i.e., the inverse Fourier transform of the visibilities), of individual bursts and combinations of bursts. All images, of the burst(s) and the corresponding PRS, were made with a cell-size of $1$\,mas and Briggs weighting with a robustness parameter\footnote{Our robustness parameter of $0.5$ in \texttt{CASA}'s \texttt{tclean} would be equivalent to a robustness parameter of $0.0$ in \texttt{AIPS}'s \texttt{imagr}.} of $0.5$.

\section{Results}\label{sec:results}

\subsection{Burst properties}
\subsubsection{Effelsberg radio telescope bursts}

\begin{figure*}[ht]
\centering
\includegraphics[width=1.0\linewidth]{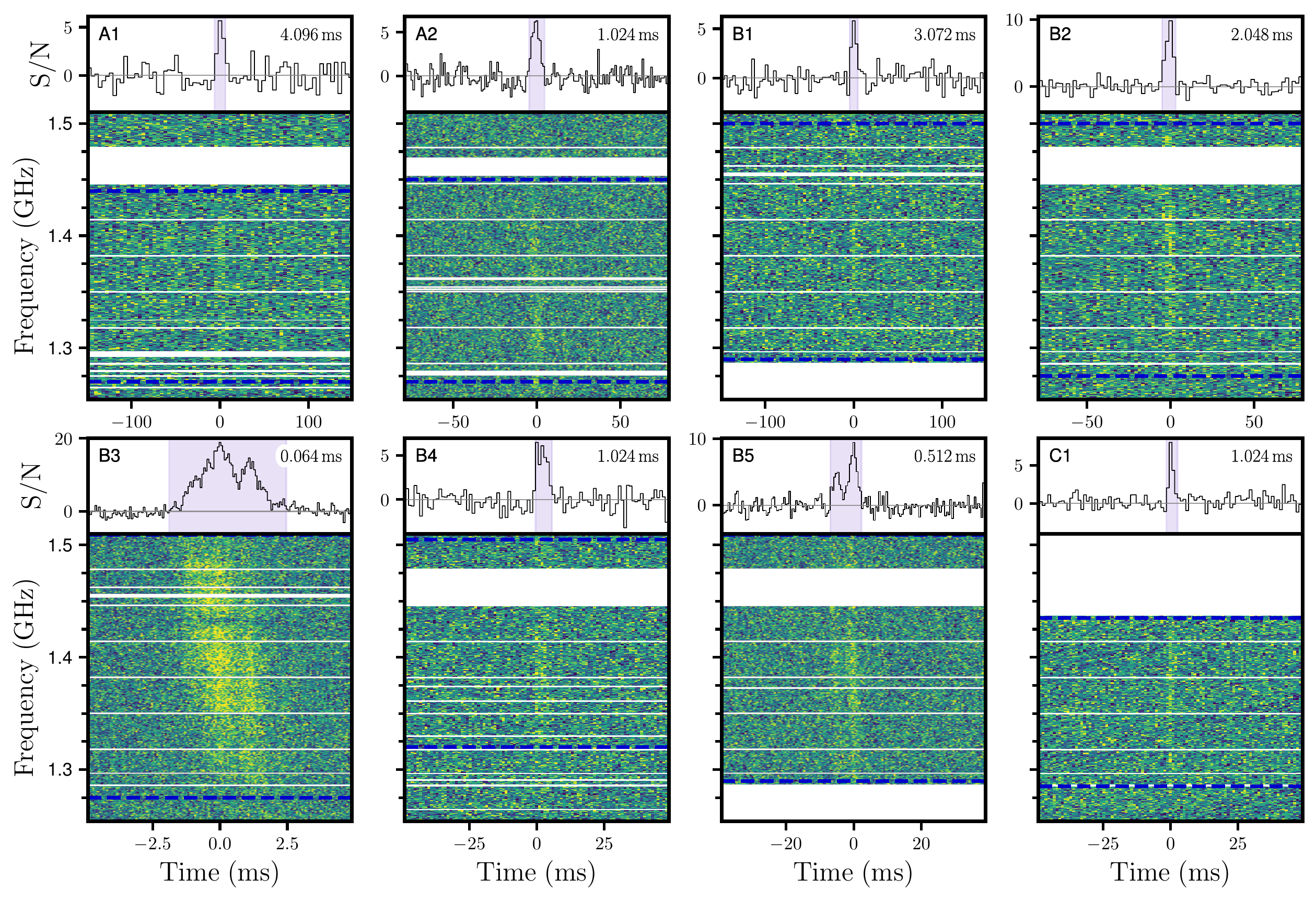}
\caption{\textbf{Temporal profiles (top sub-panels) and dynamic spectra (bottom sub-panels) of the bursts} that were detected with the Effelsberg radio telescope. Every burst is coherently de-dispersed to a DM of $551.92$\,pc\,cm$^{-3}$ and is shown with a frequency resolution of $500$\,kHz. The time resolutions used for plotting are shown in the top-right corners. Horizontal white bands are frequency channels that are flagged because of RFI or because those are the subband edges where the telescope sensitivity is significantly lower. Horizontal dashed blue lines are the manually determined frequency ranges over which the dynamic spectra are averaged to create the timeseries. The manually determined vertical regions in the timeseries indicate the start and stop times of the bursts. For visual purposes the limits of the colour map have been set to the $2^{\mathrm{nd}}$ and $98^{\mathrm{th}}$ percentile of each dynamic spectrum.}
\label{fig:family_plot}
\end{figure*}

We found $8$ bursts across three of the Effelsberg radio telescope observations. The bursts were coherently dedispersed to a DM of $551.92$\,pc\,cm$^{-3}$ (Appendix~\ref{app:dm_opt}) with \texttt{SFXC}. The resulting files contain the Stokes~I (total intensity) data with a time and frequency resolution of $32$\,$\upmu$s and $500$\,kHz, respectively, and were written out as $32$-bit filterbank files. The dynamic spectra of these bursts are shown in Figure~\ref{fig:family_plot} where they are averaged in time to increase their visibility. For every burst we determine its time of arrival (ToA), temporal width, bandwidth, peak S/N, fluence, spectral luminosity, and peak flux density (see Table~\ref{tab:burst_properties}).

\begin{table*}[ht]
    \centering
    \caption{Burst properties.}
\begin{tabular}{cclccccccc}
     \hline
     \hline
Burst & ToA$^{\text{a}}$ & PRECISE/EVN & Peak$^{\text{a}}$ & Width$^{\text{d}}$ & BW$^{\text{e}}$ & Fluence$^{\text{f}}$ & Peak Flux$^{\text{f}}$ & Spectral Luminosity$^{\text{f,g}}$ \\
label & [MJD] &  & S/N & [ms] & [MHz] & [Jy\,ms] & Density [Jy] & [$10^{31}$\,erg\,s$^{-1}$ Hz$^{-1}$] \\
     \hline
A1 & 59844.0363032356 & PR242A/EK051E           & \phantom{0}5.7 &            12.288 & 170.0 & 0.56 & 0.07 &  \phantom{0}3.57 \\
A2 & 59844.0731819791 & PR242A/EK051E           & \phantom{0}6.3 &  \phantom{0}9.216 & 180.0 & 0.80 & 0.14 &  \phantom{0}6.88 \\
B1 & 59850.0642678482 & PR243A/EK051F           & \phantom{0}5.8 &  \phantom{0}9.216 & 210.0 & 0.45 & 0.07 &  \phantom{0}3.87 \\
B2 & 59850.0888134914 & PR243A/EK051F           & \phantom{0}9.8 &  \phantom{0}8.192 & 225.0 & 0.76 & 0.15 &  \phantom{0}7.36 \\
B3 & 59850.1280616587 & PR243A/EK051F           & 18.9           &  \phantom{0}4.352 & 234.0 & 2.78 & 1.47 &            50.36 \\
B4 & 59850.2069308676 & PR243A/EK051F           & \phantom{0}6.5 &  \phantom{0}6.144 & 185.0 & 0.67 & 0.16 &  \phantom{0}8.57 \\
B5 & 59850.2128436797 & PR243A/EK051F           & \phantom{0}9.4 &  \phantom{0}9.216 & 219.0 & 1.15 & 0.29 &  \phantom{0}9.81 \\
C1 & 60022.5895413514 & PR268A$^{\text{h}}$\phantom{K051X} & \phantom{0}8.0 &  \phantom{0}4.096 & 150.0 & 0.39 & 0.19 &  \phantom{0}7.47 \\
\hline
\multicolumn{9}{l}{$^{\text{a}}$ Corrected to the Solar System Barycentre to infinite frequency assuming a dispersion measure of $551.92$\,pc\,cm$^{-3}$,} \\
       \multicolumn{9}{l}{\quad a reference frequency of $1494$\,MHz, a dispersion measure constant of $1/(2.41 \times10^{-4})$\,MHz$^{2}$\,pc$^{-1}$\,cm$^{3}$\,s and a}  \\
       \multicolumn{9}{l}{\quad source of position of \evnburstslband.}  \\
       \multicolumn{9}{l}{\quad The times quoted are dynamical times (TDB).} \\
       \multicolumn{9}{l}{$^{\text{c}}$ The peak value of the timeseries, as shown in Figure~\ref{fig:family_plot}.} \\
       \multicolumn{9}{l}{$^{\text{d}}$ Manually determined time span of the burst, shown as the highlighted region in the timeseries in Figure~\ref{fig:family_plot}.} \\
       \multicolumn{9}{l}{$^{\text{e}}$ Manually determined frequency span of the burst, i.e., the difference between the dashed cyan lines in Figure~\ref{fig:family_plot}.} \\
       \multicolumn{9}{l}{$^{\text{f}}$ We estimate a (conservative) error of 20\% for these values, which is dominated by the uncertainty in the system} \\
       \multicolumn{9}{l}{\quad equivalent flux density (SEFD) of the Effelsberg telescope ($\mathrm{T}_{\mathrm{sys}} = 20$\,K and $\mathrm{Gain} = 1.54$\,K/Jy)} \\
       \multicolumn{9}{l}{$^{\text{g}}$ Using Equation~5 from \citet{ouldboukattine_2024_arxiv} and assuming a luminosity distance of $972$\,Mpc ($z = 0.193$).} \\
       \multicolumn{9}{l}{$^{\text{h}}$ No interferometric data was produced for this observation and thus it has no EVN-code.} \\
       
    \end{tabular}
    \label{tab:burst_properties}
\end{table*}

\subsubsection{NRT bursts}
We find $18$ bursts in $125$ observations ($\sim$$106$\,hours; Appendix~\ref{app-sec:nrt_bursts_logs}). Baseband data were not recorded during the NRT observations and therefore the bursts have a best possible time and frequency resolution of $16$\,$\upmu$s and $4$\,MHz, respectively; and their intra-channel dispersion is corrected for using an MJD-dependent DM-value (Appendix~\ref{app-sec:nrt_bursts_logs}). We show the bursts and tabulate their properties in Appendix~\ref{app-sec:nrt_bursts_logs}.

\subsection{Interferometric results}
\label{subsec:loc_results}
\begin{figure*}[ht]
\centering
\includegraphics[width=1.0\linewidth]{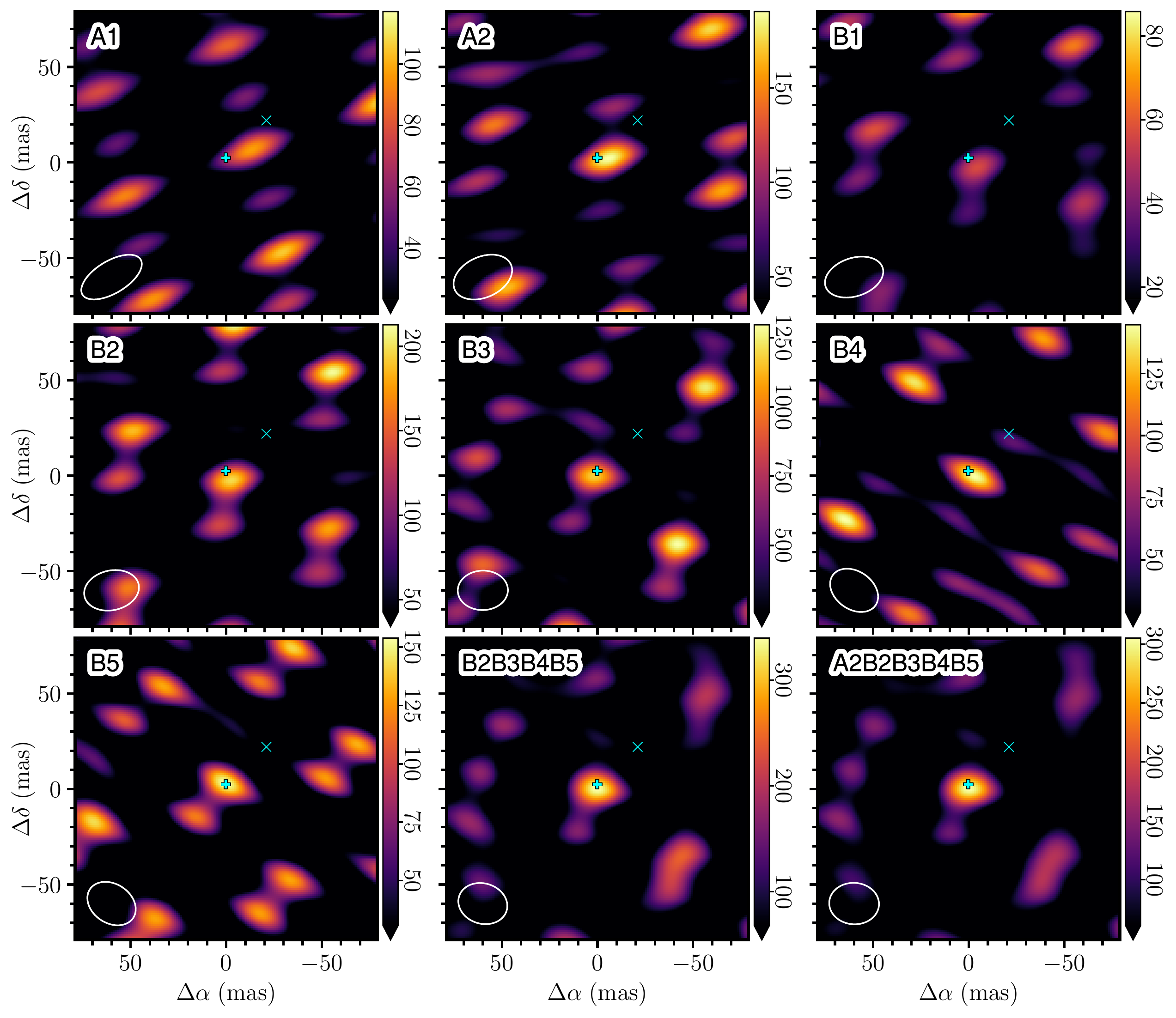}
\caption{\textbf{Dirty maps} of individual bursts and the combined visibilities of multiple bursts, as indicated by the label in the top left corner of every panel. The white ellipse in the bottom left corners show the FWHM of the synthesised beams. The colour bar is in units of mJy/beam and the limits are $20$--$100$\,\% of the maximum value (see Appendix~\ref{app:burst_loc_images} for a version without limits on the colour map). The $(0,0)$ point is \evnburstslband, which is the fitted position of the combination of five bursts (A$2$, B$2$, B$3$, B$4$, and B$5$; bottom right panel). Every panel is $80\times80$\,mas and was made with a cell-size of $1$\,mas and Briggs weighting with a robustness parameter of $0.5$ (see Appendix~\ref{app:burst_loc_images} for a version of this figure that shows $800\times800$\,mas panels). The cyan `\textbf{+}' is the C-band ($\sim$$4.5$\,GHz) PRS position from \citet{Marcote_2017_ApJL} and the cyan `$\times$' illustrates the phase reference centre.}
\label{fig:bursts_dirty_zoom}
\end{figure*}

\begin{figure}[ht]
\centering
\includegraphics[width=1.0\linewidth]{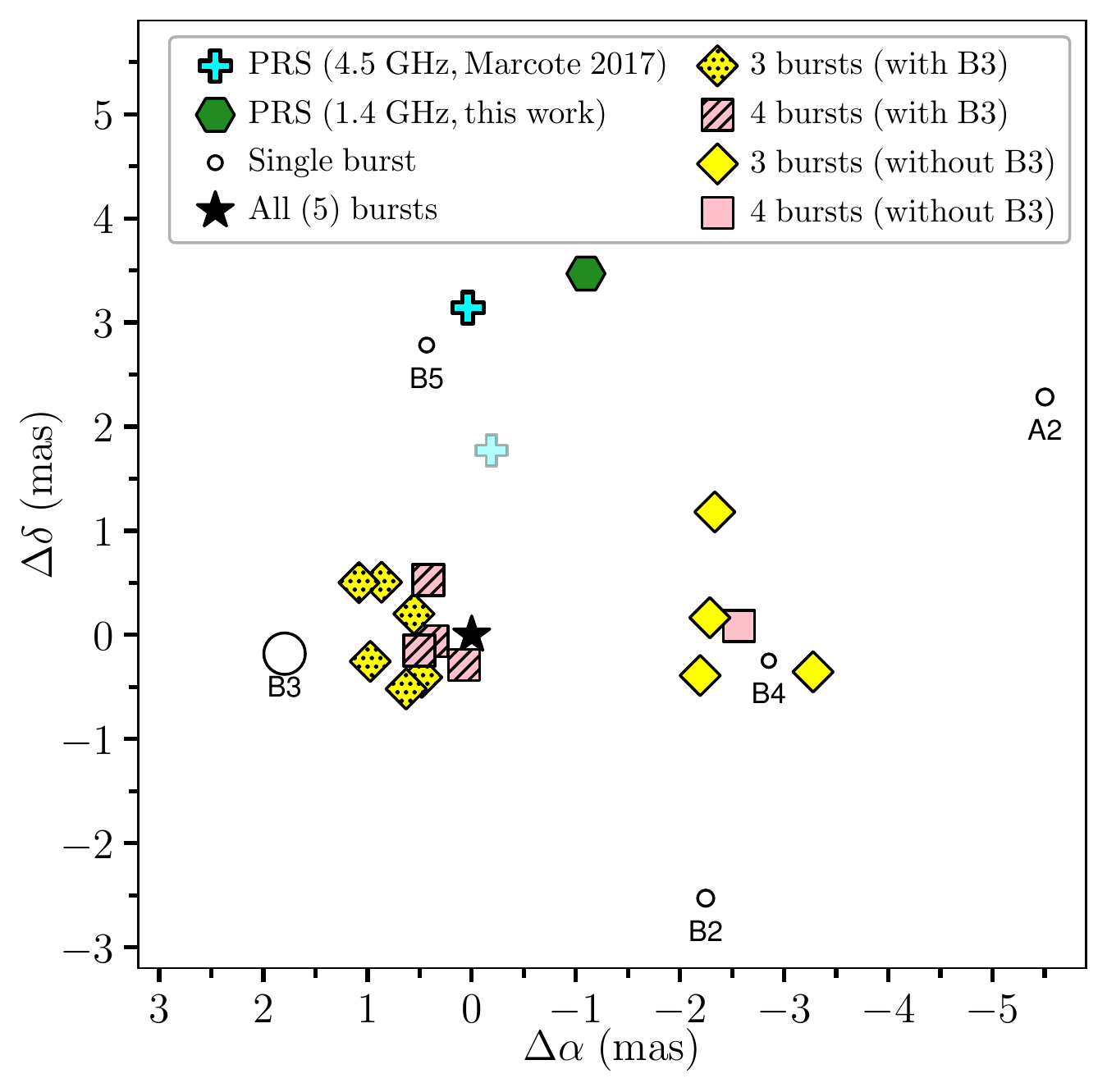}
\caption{The `best-fit' position of the bursts depends on which, and how many, bursts are used in the imaging process. The $(0,0)$ point is \evnburstslband, which is determined using bursts A2, B2, B3, B4, and B5. The green hexagon is the PRS position from this work. We show every possible combination with $3$ and $4$ bursts as yellow diamonds and pink squares. These are filled if our brightest burst, B3, is included in the imaging. Burst localisations using a single burst are illustrated with a black circle, whose sizes scale with the peak flux density of the individual burst. The solid cyan `\textbf{+}' is the Radio Fundamental Catalog (RFC)-updated C-band ($\sim$$4.5$\,GHz) PRS position from \citet{Marcote_2017_ApJL}. The transparent cyan `\textbf{+}' shows the C-band PRS position if updates to the RFC are not taken into account (Section~\ref{subsec:loc_results}). Note that precise localisations with single bursts can only be done if the `sidelobe ambiguity' is resolved. Also, note that the panel size ($\sim$$9\times9$\,mas) is much smaller than the size of the synthesised beam ($\sim$$25\times25$\,mas).} 
\label{fig:position_scatter}
\end{figure}

The dirty images of individual bursts exhibit a clear cross fringe pattern (Appendix~\ref{app:burst_loc_images}), given sufficient S/N. This is evident for all bursts, except for bursts A$1$ and B$1$. We attribute this to their low S/N (Figure~\ref{fig:family_plot}) and thus do not use bursts A$1$ and B$1$ in the localisation process. Due to the sparse $(u,v)$-coverage there are strong sidelobes when an individual burst is imaged (Appendix~\ref{app:burst_loc_images}), making it difficult to unambiguously determine the burst's position. Instead, we use multiple bursts and the Earth's rotation to fill in the $(u,v)$-plane. In Appendix~\ref{app:burst_loc_images} we show that we can solve the sidelobe ambiguity for every possible combination that uses $3$ or more bursts (excluding bursts A$1$ and B$1$). Interestingly, this also shows that for our brightest burst, B$3$, the correct lobe is not the lobe with the highest pixel value (Figure~\ref{fig:bursts_dirty_zoom}). We fit for the positions of individual bursts, as well as for combinations using $3$, $4$, or $5$ bursts, and show the results in Figure~\ref{fig:position_scatter} and Appendix~\ref{app:burst_loc_images}. Figure~\ref{fig:position_scatter} shows that the exact burst position depends on which, and how many, bursts are used in the imaging process. However, shifts in the position are on the order of $\sim$$2.5$\,mas, which is roughly $10$\,\% of the FWHM of the synthesised beam. The fitted position of the combination of five bursts (A$2$, B$2$, B$3$, B$4$ and B$5$) is: \evnburstslbanderror 

\begin{figure*}[ht]
\centering
\includegraphics[width=1.0\linewidth]{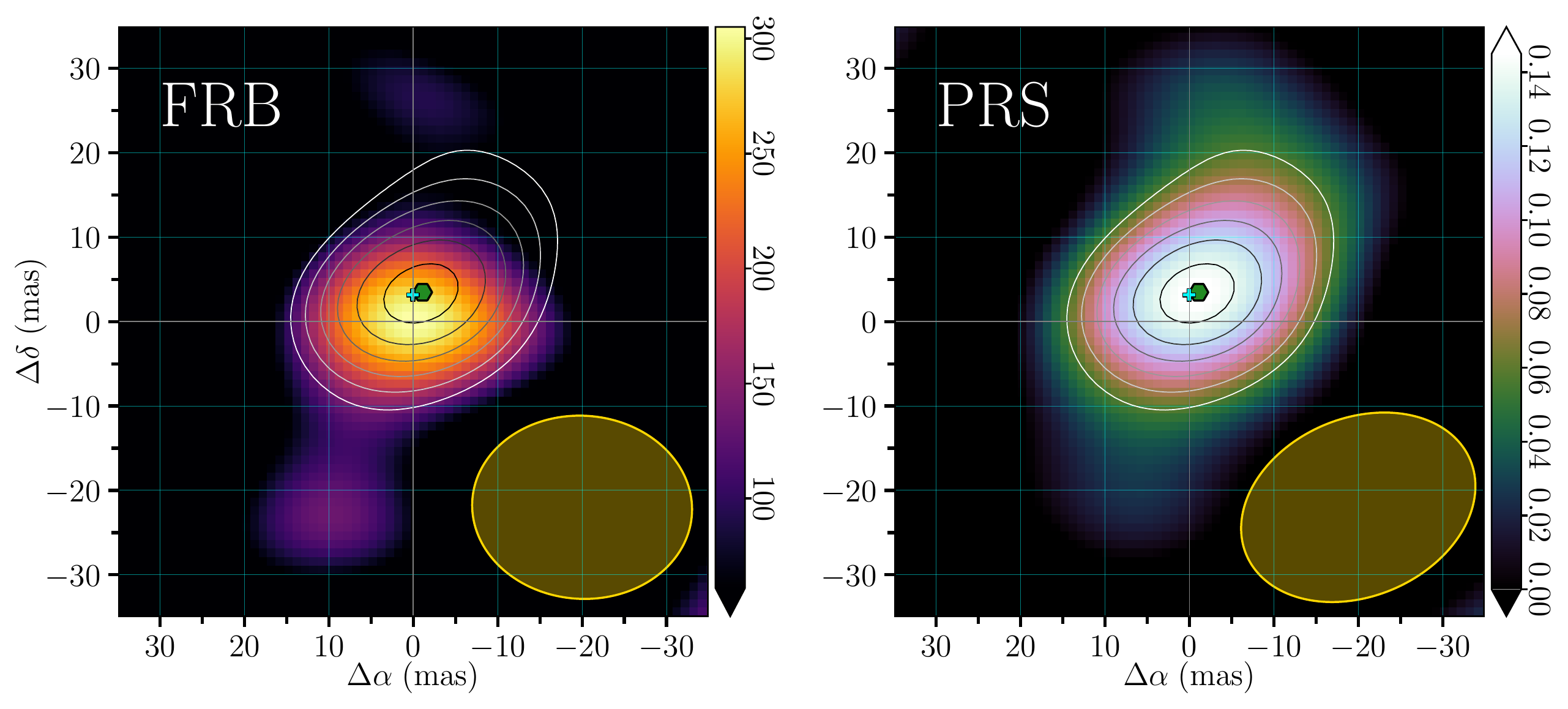}
\caption{\textbf{Left:} the dirty map of the combined visibilities of bursts A$2$, B$2$, B$3$, B$4$, and B$5$ (i.e., the bottom right panel of Figure~\ref{fig:bursts_dirty_zoom}) and the FWHM of the synthesised beam (gold ellipse). \textbf{Right:} The dirty map of the PRS (background colours) and the FWHM of the synthesised beam (gold ellipse). \textbf{Both:} we plot the contour levels of the PRS, which start at $5\sigma$, where $\sigma = 11$\,$\upmu$Jy\,beam$^{-1}$, and increase in factors of $\sqrt{2}$. In both panels the colour bar is in units of mJy/beam, the $(0,0)$ point is \evnburstslband, the cyan `\textbf{+}' is the C-band ($\sim$$4.5$\,GHz) PRS position from \citet{Marcote_2017_ApJL}, and the green hexagon is the fitted position of the PRS from this work. Both panels were made with a cell-size of $1$\,mas and Briggs weighting with a robustness parameter of $0.5$. See Appendix~\ref{app:burst_loc_images} for a zoomed-out version of the PRS without limits on the colour map.}
\label{fig:bursts_prs}
\end{figure*}

The PRS that is associated with \rone is clearly detected and shown in Figures~\ref{fig:bursts_prs} and Appendix~\ref{app:burst_loc_images}. In Figure~\ref{fig:bursts_prs} contour lines of the PRS flux are drawn, which start at five times the standard deviation of the outer regions of the dirty image (thus excluding the PRS itself) and increase by factors of $\sqrt{2}$. The flux density of the PRS is $S_{{1.4\,\mathrm{GHz}}} = 155 \pm 26$\,$\upmu$Jy\,beam$^{-1}$. The uncertainty on the flux density is the combination of the standard deviation of the dirty image and a $15$\,\% calibration error\footnote{\url{https://www.evlbi.org/evn-data-reduction-guide}} added in quadrature. Similar to \citet{moroianu_2025_arxiv}, we do not remove the burst windows from the continuum data. Given the fluences of the bursts (Table~\ref{tab:burst_properties}) and the duration of the observations (Appendix~\ref{app:obs_timeline}), we estimate that the bursts contribute $\lesssim$$0.2$\,$\upmu$Jy to the continuum data. Since this is roughly two orders of magnitude less than the thermal noise in our images, this contribution can be dismissed. We calculate the luminosity as:
\begin{equation}
    L_{\nu} = 4 \pi D_{L}^{2} S_{\nu} \left( 1 + z \right)
\end{equation}
Assuming a luminosity distance, $D_{L}$, to \rone of $972$\,Mpc, isotropic emission and $z = 0.193$ \citep{tendulkar_2017_apjl}, this corresponds to a luminosity of $L_{1.4\,\mathrm{GHz}} = (2.09 \pm 0.35) \times 10^{29}$\,erg\,s$^{-1}$\,Hz$^{-1}$. The fitted position of the PRS, combining both epochs is: \evnprslbanderror The separation between the position of the bursts and the position of the PRS is $3.6$\,mas, which, given the distance\footnote{Note that the luminosity distance is not the same as the angular diameter distance. At a redshift of $z = 0.193$ \citep{tendulkar_2017_apjl}, and adopting cosmological parameters from \citet{planck_2020_aa}, we find a transverse scale size of $3.31$\,pc$/$mas.} to \rone, translates to $\sim$$12$\,pc. Given that the sizes of the synthesised beams of the burst data and the PRS data are roughly equal, we also estimate the error on the PRS position to be $\sim$$2.5$\,mas. Therefore, the bursts and the PRS are $1\sigma$ consistent with being co-located.\\

To compare these positions with the previously determined C-band ($\sim$$4.5$\,GHz) PRS detection \citep{Marcote_2017_ApJL}, updates to the Radio Fundamental Catalog \citep[RFC;][]{petrov_2025_apjs} need to be taken into account. In both this work, as in \citet{Marcote_2017_ApJL}, the source J0529+3209 was used as the phase calibrator. The assumed position of this source has shifted by $+0.3$\,mas in right ascension and by $+1.4$\,mas in declination between the RFC versions of 2016 \citep{Marcote_2017_ApJL} and 2022 (this work). Thus, to make the comparison between 2016 and 2022 correct, we shift the C-band PRS position with the same increase in right ascension and declination. In all applicable figures, the updated C-band PRS position is shown as a solid solid cyan `\textbf{+}'. In Figure~\ref{fig:position_scatter} we also show the C-band PRS position if updates to the RFC are not taken into account. 

\subsection{DM evolution}
\label{sec:dm_evolution}
\begin{figure*}[ht]
\centering
\includegraphics[width=1.0\linewidth]{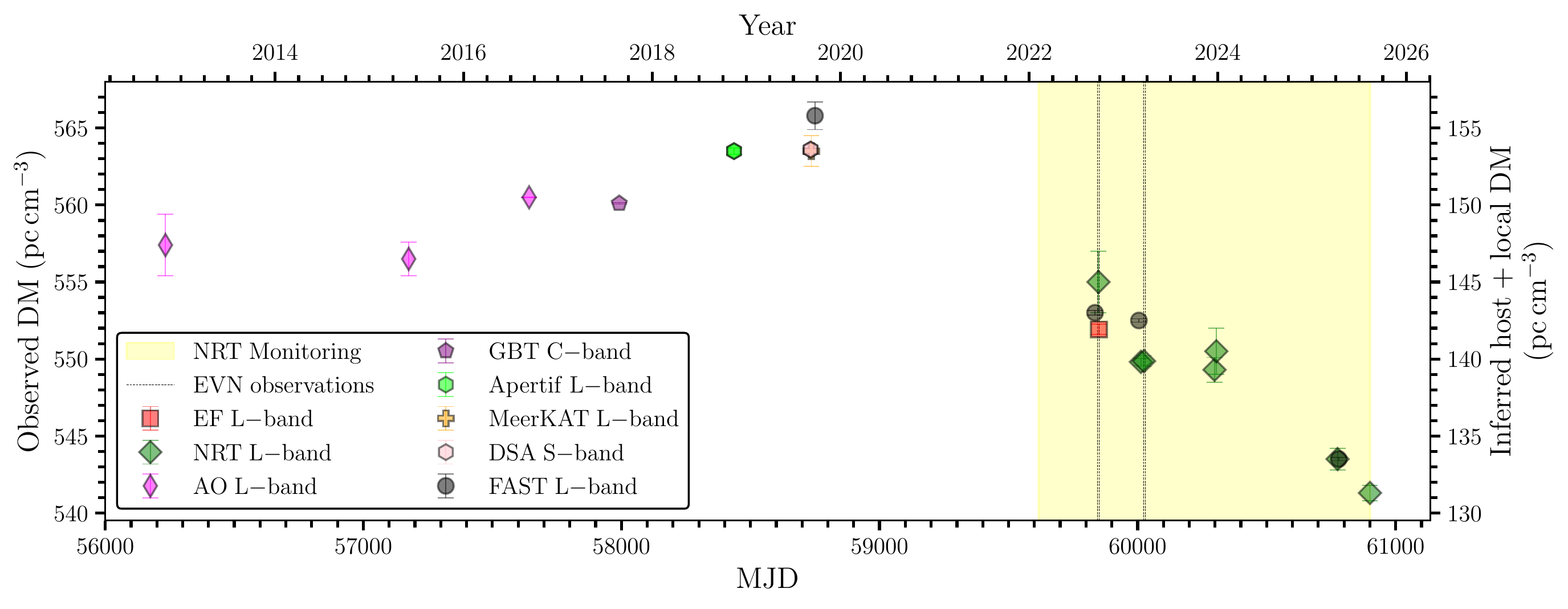}
\caption{\textbf{Dispersion measure} as a function of time. This figure shows the DM measurements from this work: Burst B$3$ (EF, Effelsberg telescope, red square), along with various bursts from the Nan\c{c}ay radio telescope (NRT, green diamonds). Other previously determined DM values are also shown and include bursts from the Arecibo Observatory \citep[AO][]{spitler_2014_apjl,spitler_2016_nature,hessels_2019_apjl}, Green Bank Telescope \citep[GBT;][]{snelders_2023_natas}, Aperture Tile in Focus \citep[Apertif;][]{oostrum_2020_aa}, MeerKAT \citep{platts_2021_mnras}, Deep Space Network \citep[DSN;][]{majid_2020_apjl}, and the Five-hundred-meter Aperture Spherical radio Telescope \citep[FAST;][]{li_2021_nature,wang_2025_arxiv,zhang_2025_atel}. The inferred host $+$ local DM is the observed DM minus the expected DM contributions of the Milky Way and intergalactic medium (Section~\ref{sec:dm_evolution}).}
\label{fig:dm_evolution}
\end{figure*}

We determine the best structure-maximized DM for burst B$3$ (Appendix~\ref{app:dm_opt}) and find that it is $551.92 \pm 0.33$\,pc\,cm$^{-3}$. We do not determine the DM for the other bursts found with the Effelsberg radio telescope due to their low S/N (Figure~\ref{fig:family_plot}). For bursts detected with NRT, we determine the DM for the brightest burst of every MJD, which are tabulated in Table~\ref{tab:burst_properties_NRT}. These values, together with previously published DM measurements of \rone, are plotted as a function of time in Figure~\ref{fig:dm_evolution}. Since \rone's discovery, the DM first increased from $\sim$$560$\,pc\,cm$^{-3}$ to $\sim$$566$\,pc\,cm$^{-3}$ (late $2019$), and then a substantial decrease of $\sim$$25$ units was observed in $\sim$$5$\,years time (DM of $\sim$$541$\,pc\,cm$^{-3}$ in August 2025).

To determine the fractional change in the DM contribution of the host galaxy and the local environment, DM$_{\textrm{host$+$local}}$, of \rone we need to subtract the DM contributions of the Milky Way interstellar medium ($\mathrm{DM}_{\mathrm{MW,ISM}}$), the Milky Way halo ($\mathrm{DM}_{\mathrm{MW,Halo}}$), and the intergalactic medium ($\mathrm{DM}_{\mathrm{IGM}}$) from the observed DM. The Galactic electron density model \texttt{NE2001p} estimates $\mathrm{DM}_{\mathrm{MW,ISM}} = 188$\,pc\,cm$^{-3}$ for the line of sight (LoS) of \rone \citep{ocker_2024_rnaas}. However, two new high-DM pulsars near the LoS of \rone ($l = 174.88\degr$, $b = -0.22\degr$) were recently discovered; PSR~J0557+2442g ($l = 184.97\degr$, $b = +0.08\degr$) is $10\degr$ away from \rone has a DM of $195$\,pc\,cm$^{-3}$ and PSR~J0517+3436g ($l = 172.04\degr$, $b = -1.82\degr$) is $3\degr$ away and has a DM of $192$\,pc\,cm$^{-3}$ \citep{han_2025_raa}. Therefore we assume a $\mathrm{DM}_{\mathrm{MW,ISM}} = 200$\,pc\,cm$^{-3}$ for the LoS of \rone. We assume $\mathrm{DM}_{\mathrm{MW,Halo}} = 41$\,pc\,cm$^{-3}$ \citep{yamasaki_2020_apj}, which is supported by \citet{cook_2023_apj} constraints placed at higher Galactic latitudes. Finally, we assume $\mathrm{DM}_{\mathrm{IGM}} = 169$\,pc\,cm$^{-3}$ \citep[using the relation from][with $z = 0.193$]{macquart_2020_nature}. The inferred DM of the host galaxy and the local environment of \rone are shown on the right side of Figure~\ref{fig:dm_evolution}. We assume that $\mathrm{DM}_{\mathrm{host}}$ remains constant and thus show that the fractional change in the DM of the local environment, $\mathrm{DM}_{\mathrm{local}}$, is $\gtrsim$$15$\,\%. 

\subsection{Polarimetry}
\label{subsec:pol}
We do not find any significant linear or circular polarisation in our brightest burst, B$3$. Circular polarisation in bursts from \rone occur in less than $1$\,\% of the bursts \citep{feng_2022_sciencebulletin} and it is known that the linear polarisation fraction decreases with decreasing frequency \citep[e.g.,][]{Plavin_2022_MNRAS}. Based on \citet{wang_2025_arxiv} we suspect that at our observing frequencies ($\sim$$1.4$\,GHz) the linear polarisation fraction is $\sim$$5$\,\% and that we do not have enough S/N in B$3$ to measure any significant linear polarisation. We do not investigate the NRT data for any polarimetric properties due to the large intra-channel depolarisation (Appendix~\ref{app-sec:pol}) and the other Effelsberg-detected bursts are not investigated due to their low S/N.

\section{Discussion}\label{sec:discussion}
In this work, we used the EVN to strongly constrain the offset between the \rone burst source and its associated PRS, demonstrating that they are co-located to within $\lesssim$$ 4$\,mas, which is equivalent to $\lesssim$$12$\,pc transverse offset in the host galaxy. This analysis significantly improves on the previous work of \citet{Marcote_2017_ApJL}, where limited $(u,v)$-coverage led to a 4$\times$ larger uncertainty in the position of the \rone burst source. In addition, we find that our PRS flux density measurements are consistent with other observations, taken at different epochs and with varying spatial resolution \citep[e.g.,][]{rhodes_2023_mnras}; and, we map the time-variable DM of \rone, combining our data with measurements from many other telescopes \citep{spitler_2014_apjl,spitler_2016_nature,hessels_2019_apjl,snelders_2023_natas,oostrum_2020_aa,platts_2021_mnras,majid_2020_apjl,li_2021_nature,wang_2025_arxiv,zhang_2025_atel}. \rone is one of the best-studied FRB sources, with a wealth of observational information collected over the past decade (see Introduction and references therein). Building on these studies and the work we present here, we discuss four possible astrophysical scenarios for its nature:
\\
\\
\noindent1) \textit{Isolated magnetar}: In this scenario, the \rone burst source is an isolated magnetar and the PRS is a rotation-powered pulsar wind nebula (PWN) or magnetically powered magnetar wind nebula \citep[MWN;][]{margalit_2018_mnras,rahaman_2025_apj}. Alternatively, or in addition to the PWN/MWN scenario, the PRS could also represent the remnant from a super-luminous supernova (SLSN) or a long gamma-ray burst (LGRB) that is interacting with the circumstellar medium. The bursts are created by explosive magnetic reconnection events \citep{lyubarsky_2020_apj}, curvature radiation \citep{kumar_2017_mnras}, a synchrotron maser \citep[e.g.,][]{lyubarsky_2014_mnras,beloborodov_2017_apjl,metzger_2019_mnras}, or some other emission mechanism \citep[e.g.,][]{thompson_2019_apj,long_2023_prd}. The PRS is powered by sporadic magnetar flares and/or could result from magnetic braking of a magnetar that was born spinning rapidly (period of tens of milliseconds or less) and quickly spun-down, dumping its kinetic energy into its surroundings. The closest-known Galactic analogues would be SGR~1935+2154, which created much weaker but still FRB-like bursts \citep{bochenek_2020_nature,chime_2020_nature_sgr,kirsten_2021_natas}; the Crab Nebula, though it is $\sim$$10^{4}$--$10^{5}$$\,\times$ less luminous than the \rone PRS \citep[e.g.,][and references therein]{delooze_2019_mnras}; and Swift~J1834.9$-$0846, the only Galactic magnetar with a known nebula \citep{younes_2016_apj}.
\\
\\
\noindent2) \textit{Accreting binary}: In this scenario, \rone is a black hole (BH) or neutron star (NS) that is accreting at hyper-Eddington rates from a binary stellar companion and launching a relativistic jet into the circum-binary medium. The PRS would be the synchrotron radiation from electrons energized at the disk wind/jet termination shock \citep{sridhar_2022_apj,sridhar_2024_apj}, and the bursts would be generated via reconnection of striped magnetic fields inside the jet or due to magnetized forward shocks harbouring synchrotron maser instability \citep{Sridhar_2021_ApJ}. In this scenario, the bursts are beamed along the jet's narrow funnel. The jetted Galactic microquasar SS~433 and its SNR W~50 nebula provide a potential analogue, though the accretion rate of this system is uncertain; furthermore, the precessing and flaring jet associated with SS~433 is not aligned to our line of sight, which would prevent the detection of any FRBs. It is thought that SS~433 could be an off-axis Galactic equivalent of ultra-luminous X-ray sources (ULXs). Extragalactic ULXs provide an example of hyper-Eddington accretors; however, none are known to have produced an FRB-like burst, though potentially due to a lack of concerted search for ms-duration flares from known highly-accreting `microblazars' --- i.e., stellar-mass, accreting compact objects with a jet pointed at us.
\\
\\
\noindent3) \textit{Interacting binary}: An alternative binary scenario involves magnetic interaction and/or wind interaction between two closely orbiting bodies, one of which could be a magnetar or at least a high-B-field neutron star \citep{zhang_2017_apjl,mottez_2020_aap}. The recent discovery that some long-period transients are magnetically interacting white dwarf -- M-dwarf binaries \citep{deruiter_2025_natas} provides a potential analogy, though the mechanism would need to scale-up enormously in energy to explain \rone's bursts. In this scenario, the radio bursts could be from particles accelerated along a magnetic bridge connecting the two bodies; the origin of the PRS would be from an intra-binary shock between the stellar winds \citep[e.g.,][and references therein]{marcote_2016_mnras}.
\\
\\
\noindent4) \textit{Indirect association}: Though the \rone burst source and the PRS are extremely close to each other, it remains possible that they are not physically associated, at least not in a direct sense. In this scenario, the PRS could be a `wandering' or `off-nuclear' massive accreting BH \citep[MBH;][]{reines_2020_apj}. Low-Eddington accretion onto an MBH, with a mass of $10^{4}$--$10^{7}$\,$\mathrm{M}_{\odot}$, would power a (near-)constant luminosity PRS through small-scale jets. The source of the FRBs would be a compact object, most likely a magnetar, that resides within $0.1$--$10$\,pc of the MBH and happens to be there because the MBH is embedded in a star cluster \citep{askar_2023_arxiv}. An analogue would be the Galactic Centre magnetar SGR~J1745$-$2900, which is situated $\sim$$0.1$\,pc from Sgr~A$^{\mathrm{*}}$ \citep{desvignes_2018_apjl}.

\subsection{Connecting observables to proposed scenarios}

\noindent\textit{PRS luminosity and stability}: Over a decade of monitoring, there is no clear, secular evolution of the \rone luminosity. Nor is there evidence that the flux is significantly resolved on milliarcsecond scales compared to lower-resolution observations \citep[e.g.,][]{Marcote_2017_ApJL,rhodes_2023_mnras}. Stochastic luminosity variations at the tens of percent level are likely due to refractive scintillation in the foreground interstellar medium of the Milky Way \citep{bhardwaj_2025_arxiv}. The stable luminosity of the PRS disfavours a supernova interaction scenario, and is consistent with a relatively continuous injection of energy into an expanding PWN/MWN; or, outflows in the form of accretion disk winds or a relativistic jet.
\\
\\
\noindent\textit{Compactness and co-location}: The strict co-location between the \rone burst source and PRS is consistent with all the proposed scenarios. Furthermore, previous VLBI observations at 5\,GHz have shown that the PRS must have a physical size that is smaller than $0.7$\,parsec in diameter \citep{Marcote_2017_ApJL}. For comparison, the Crab nebula is $\sim$$1,000$\,years old and has a diameter of $\sim$$4$\,pc \citep[assuming an angular diameter of $7.5\arcmin$ and a distance to the Crab of $1.9$\,kpc,][]{lin_2023_apj}. The \rone-system could be similar to the Crab pulsar/nebula, but at a much younger age with a still expanding nebula. Alternatively, if \rone would be orbiting an MBH at a period of $\sim$$160$\,days \citep{braga_2025_arxiv}, then the separation between the burst source and the MBH would be a mere $10$--$100$\,AU (for BH masses of $10^{4}$--$10^{7}$\,$\mathrm{M}_{\odot}$) --- which will never be resolved by Earth-based VLBI, given the distance to \rone. Higher-resolution VLBI can, however, further constrain the source size to the point of detecting or ruling out a compact relativistic jet powered by accretion.
\\
\\
\noindent\textit{DM and RM evolution}: More so than any other known FRB, \rone showcases \textit{both} considerable RM and DM evolution, with a $\sim$$75$\,\% decrease in RM \citep[e.g.,][]{Michilli_2018_Nature,Plavin_2022_MNRAS} and a decrease of at least $\sim$$15$\,\% in $\mathrm{DM}_\mathrm{local}$ \citep[Figure~\ref{fig:dm_evolution} and][]{wang_2025_arxiv}. These changes must be occurring local to \rone because our line of sight through the intervening magneto-ionised media does not appreciably change. The burst source is clearly in an extreme and dynamic magneto-ionic environment, which also leads to depolarisation of the bursts towards lower radio frequencies \citep[e.g.,][]{Plavin_2022_MNRAS,feng_2023_atel}. DM and RM variations could point to a binary scenario, though the timescales do not match the proposed 
$\sim$$160$-day burst activity period. In the case of the pulsar binary PSR~B1259$-$63, DM and RM variations are similar in magnitude to \rone, but also clearly track with orbital phase and increase around the time of periastron \citep{johnston_2005_mnras}. In the case of the Galactic Centre magnetar, the increase in RM is an order-of-magnitude smaller than for \rone (over similar years-long timescales), and is thought to be dominated by variations in the projected magnetic field, because the DM is consistent with being constant \citep[to within $2\sigma$;][]{desvignes_2018_apjl}. 

Notably, the time evolution of \rone's DM exhibits an increasing trend followed by a decrease. The increase in DM can be explained by a scenario where the FRB engine is embedded within an SNR that sweeps up material from upstream, with mass comparable to the ejecta \citep{Piro&Gaensler_2018}. However, this scenario would also predict a steady decrease in the DM (due to the ionizing radiation from the reverse shock and the decreasing upstream density) to precede a constant/late-time DM increase phase. The observed trend challenges this picture. On the other hand, in the hypernebula scenario, \citet{sridhar_2022_apj} predicted that the contribution of the local environment (due to the nebular electrons; not the external shell) to the DM would also be non-monotonic (see their right panel of Figure~9 or Figure~6 for `toy model' parameters). Here, the local DM can gradually increase and then decrease more rapidly (depending on the engine's age, among other properties). The moderate increase in the DM would be due to the deceleration of the outflowing material after its short free-expansion phase. For \rone, this transition from increasing to decreasing DM was predicted to occur when its RM is $\sim{\rm few}\times10^4\,{\rm rad\,m^{-2}}$, as corroborated by our observations. Future, long-term monitoring would reveal whether the decrease in DM is due to a late-time reduction in the accretion rate of the system or due to stochastic fluctuations of the density/ionization fraction in the outflows. A detailed calculation of the hypernebula's parameters that fit the observed trend in DM and RM is left for future work.
\\
\\
\noindent\textit{Evolution of burst activity}: Since the first \rone burst was detected $13$ years ago \citep{spitler_2014_apjl}, the source has shown months-long periods of activity with the burst rate varying on timescales as short as tens of minutes \citep[e.g.,][]{gajjar_2018_apj,hewitt_2022_mnras}. As yet, there is no evidence for a secular decrease in \rone's activity or average burst energy \citep[e.g.,][]{wang_2025_arxiv}. \rone's proposed activity period of $\sim$$160$\,days (still to be confirmed robustly by continued monitoring) could be linked to precession, rotation or orbital motion of the burst source. It is unlikely that the period represents the rotation period of a compact object. White dwarfs have typical rotation period of $<$$4$\,days \citep[see, e.g.,][]{hernandez_2024_mnras}. While magnetars spin slower than pulsars, bridging the gap between the $7$-hr (candidate) magnetar \citep{deluca_2006_science} to the $\sim$$160$\,days period of \rone seems implausible.
\\
\\
\noindent\textit{Burst temporal scales and spectra}: \rone has only been detected once below $1$\,GHz radio frequency \citep{josephy_2019_apjl}, but this lack of activity does not necessarily stem from absorption and propagation effects in its local environment. While the majority of \rone's bursts have millisecond durations, some have been shown to last only microseconds \citep{snelders_2023_natas}. This is consistent with a small emission region size in a neutron star magnetosphere and in tension with models invoking large-scale shocks from magnetars or accreting binary scenario that require the FRBs to be produced at much larger distance from the central engine. However, if the emission is due to the merger of plasmoids resulting from magnetic reconnection, then the emission duration is determined by the size of the plasmoids and not by the emission radius. Furthermore, there are several plasma instabilities (e.g., filamentation and modulation) associated with the propagation of strong electromagnetic waves that can imprint fine spectro-temporal-structures onto the pulse \citep[e.g.,][]{sobacchi_2021_mnras}.

\section{Future observations}
While we cannot strictly rule out that \rone and its coincident PRS are only indirectly associated, the existence of now several FRB-PRS systems \citep{moroianu_2025_arxiv}, in dwarf galaxies, with high and variable local DM/RM, suggests they could be a distinct type of FRB source and not just a matter of circumstance. We identify the following as the most promising ways to clarify the nature of \rone and its PRS:

\begin{itemize}

\item Global VLBI observations at frequencies $\gtrsim$$8$\,GHz (X-band) will place even stronger constraints (expected to be $\sim$$2$ times better than \citealt{Marcote_2017_ApJL}) on the size of the PRS, and may even begin to resolve it. If the PRS is resolved, these observations will also tell how fast it is expanding. A direct measurement of the PRS size could argue for the accreting binary scenario.

\item Likewise, high-sensitivity, global VLBI observations at $\gtrsim$$4$\,GHz, that detect multiple bursts from \rone could place $\sim$$3$--$10$\,$\times$ stronger constraints on the co-location with the PRS. Any measurable separation would argue for the indirect association scenario. Furthermore, these measurements will place constraints on the burst source's kick velocity if the PRS is an SNR.

\item If the luminosity of the PRS is slowly decreasing --- e.g., due to rotational slowdown or decreased accretion rate --- it might take a couple decades before this can be robustly disentangled from the variability of the PRS due to interstellar scintillation. For context, the flux density of the Cassiopeia~A SNR, estimated to be about $350$ years old, decreases by $\sim$$0.7$\,\% per year \citep[see, e.g.,][]{trotter_2017_mnras}.

\item Currently the lowest-frequency detection of the PRS is at $400$\,MHz \citep{resmi_2021_aa}. The spectral energy distribution (SED) of the PRS is nearly flat between $400$\,GHz and $10$\,GHz \citep[see, e.g., Figure~2 of][]{resmi_2021_aa}. Since \rone is located in the direction of the Galactic plane ($l = 174.88\degr, b = -0.22\degr$), interferometric observations at low frequencies are challenging. Nonetheless, observations at $\sim$$150$\,MHz with, e.g., LOFAR would be useful to see if the PRS is affected by self-absorption. Likewise, it would be interesting to see if, and how, the shape of the SED changes over time, since the spectral index is not expected to be constant over the course of a MWN or hypernebula's lifetime.

\item Lastly, we encourage continued monitoring of \rone, also outside of its expected activity window, to check for DM and RM evolution and periodic activity. A prolonged DM decrease towards \rone could be explained in an expanding SNR scenario. Likewise, a sign flip in the RM \citep{annathomas_2023_science} of \rone could suggest that the system is in a binary system. Also, changes in the scattering and/or scintillation properties of the bursts would be interesting, since they will largely track changes in the local environment.

\end{itemize}

\clearpage

\begin{acknowledgements}
We thank Ziggy Pleunis and Amanda Cook for useful discussions.
\\
\indent
The European VLBI Network is a joint facility of independent European, African, Asian, and North American radio astronomy institutes. Scientific results from data presented in this publication are derived from the following EVN project code(s): EK051. 
The Nan\c{c}ay Radio Observatory is operated by the Paris Observatory, associated with the French {\it Centre National de la Recherche Scientifique} (CNRS). We acknowledge financial support from the {\it Programme National de Cosmologie et Galaxies} (PNCG) and {\it Programme National Hautes Energies} (PNHE) of INSU, CNRS, France.
This work is based on observations with the $100$-m telescope of the MPIfR (Max-Planck-Institut f\"ur Radioastronomie) at Effelsberg (Germany).
This work is based in part on observations carried out using the 32-m radio telescope operated by the Institute of Astronomy of the Nicolaus Copernicus University in Toru\'n (Poland) and supported by a Polish Ministry of Science and Higher Education SpUB grant.
\\
\indent
The AstroFlash research group at McGill University, University of Amsterdam, ASTRON, and JIVE is supported by: a Canada Excellence Research Chair in Transient Astrophysics (CERC-2022-00009); an Advanced Grant from the European Research Council (ERC) under the European Union’s Horizon 2020 research and innovation programme (`EuroFlash'; Grant agreement No. 101098079); and an NWO-Vici grant (`AstroFlash'; VI.C.192.045).
N.S. acknowledges support from the Simons Foundation (grant MP-SCMPS-00001470).
B.M. acknowledges financial support from the State Agency for Research of the Spanish Ministry of Science and Innovation, and FEDER, UE, under grant PID2022-136828NB-C41/MICIU/AEI/10.13039/501100011033, and through the Unit of Excellence Mar\'ia de Maeztu 2020--2023 award to the Institute of Cosmos Sciences (CEX2019- 000918-M).
F.K. acknowledges support from Onsala Space Observatory for the provisioning of its facilities/observational support. The Onsala Space Observatory national research infrastructure is funded through Swedish Research Council grant No 2017-00648.
S.B. is supported by a Dutch Research Council (NWO) Veni Fellowship (VI.Veni.212.058).
K.N. is an MIT Kavli Fellow.\\

\hfill \\
\textit{Data availability:} Uncalibrated visibilities of \rone (both the bursts and the associated PRS) and its calibration sources can be downloaded from the JIVE/EVN archive, \url{https://archive.jive.eu}, under project codes EK051E and EK051F. Calibrated burst visibilities, dirty maps fits files, and the scripts that generated Figures~\ref{fig:bursts_dirty_zoom}, \ref{fig:bursts_prs}, \ref{fig:position_scatter}, \ref{fig:dm_evolution}, \ref{app-fig:obs_timeline}, \ref{app-fig:prs_zoom_out}, \ref{app-fig:burst_dirty_all}, \ref{app-fig:burst_dirty_no_cap_zoom}, \ref{app-fig:burst_dirty_no_cap}, \ref{app-fig:burst_combinations}, \ref{app-fig:no_sidelobe_ambiguity}, \ref{app-fig:depol} and Table~\ref{tab:nancay_obs_log} can be accessed in our Zenodo reproduction package: {\it Zenodo DOI will be made public once the paper is accepted}. Due to the large file sizes, the burst filterbank files and voltage data, and calibrated continuum visibilities will be made available by the authors upon reasonable request.

\end{acknowledgements}

\appendix

\section{Observational timeline}
\label{app:obs_timeline}

\begin{figure*}[ht]
\centering
\includegraphics[width=1.0\linewidth]{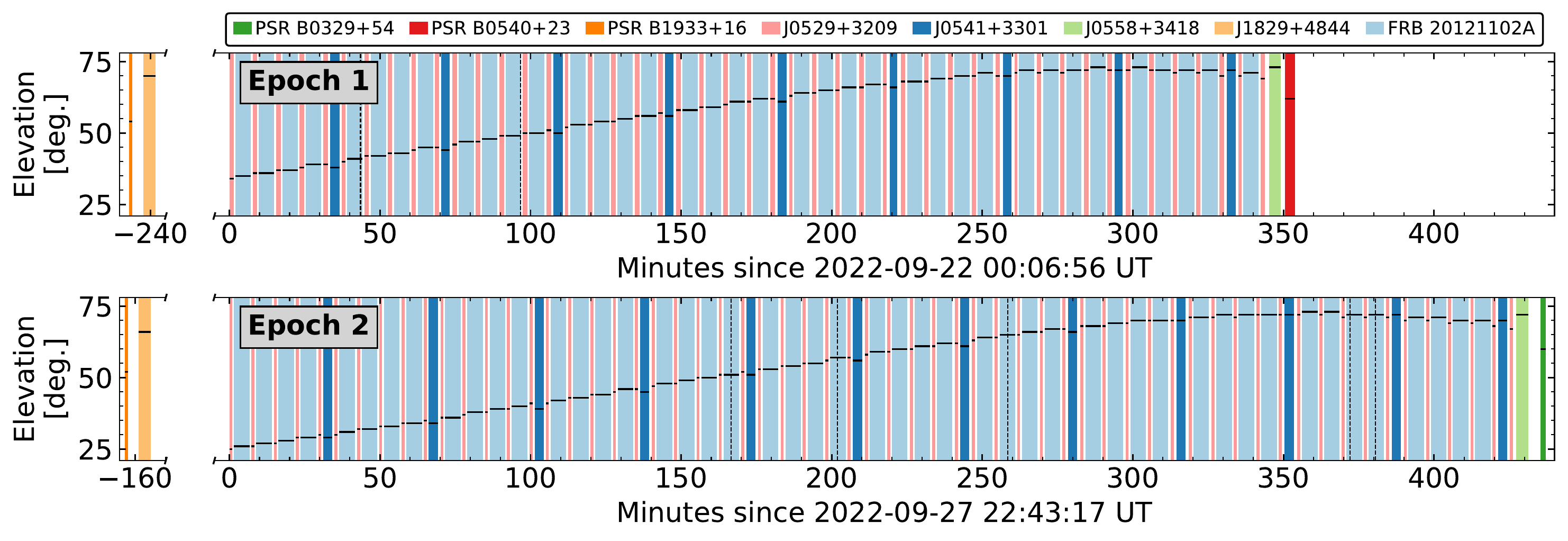}
\caption{\textbf{Observational timeline and calibration strategy} for Epoch~1 (top panel; PR242A/EK051E) and Epoch~2 (bottom panel; PR243A/EK051E). The $\sim$$6$-min target scans on \rone are interleaved with $\sim$$2$-min scans on the phase calibrator J2125+0441. Twice per hour the check source, J0541+3301, was observed for $\sim$$3$\,min per scan to verify the (astrometric) calibration. J1829+4844 was used as a fringe finder and bandpass calibrator. J0558+3418 was used for the same reasons as J1829+4844, but was not used since J1829+4844 is brighter. The pulsars B0329+54, B0540+23, and B1933+16 were used to test the data quality, frequency setup and the burst-search pipeline. The solid horizontal black bars indicate the source elevation for the Effelsberg telescope (Germany). Vertical black dashed lines indicate the arrival times of the bursts. Vertical white bars
indicate that the telescopes are slewing (slewing/on-source times are shown for Effelsberg, which is the largest and slowest-slewing antenna
in the array).}
\label{app-fig:obs_timeline}
\end{figure*}

Figure~\ref{app-fig:obs_timeline} shows the calibration strategy of our PRECISE/EVN observations.

\section{NRT bursts and observations log}
\label{app-sec:nrt_bursts_logs}

\begin{figure*}[ht]
\centering
\includegraphics[width=0.9\linewidth]{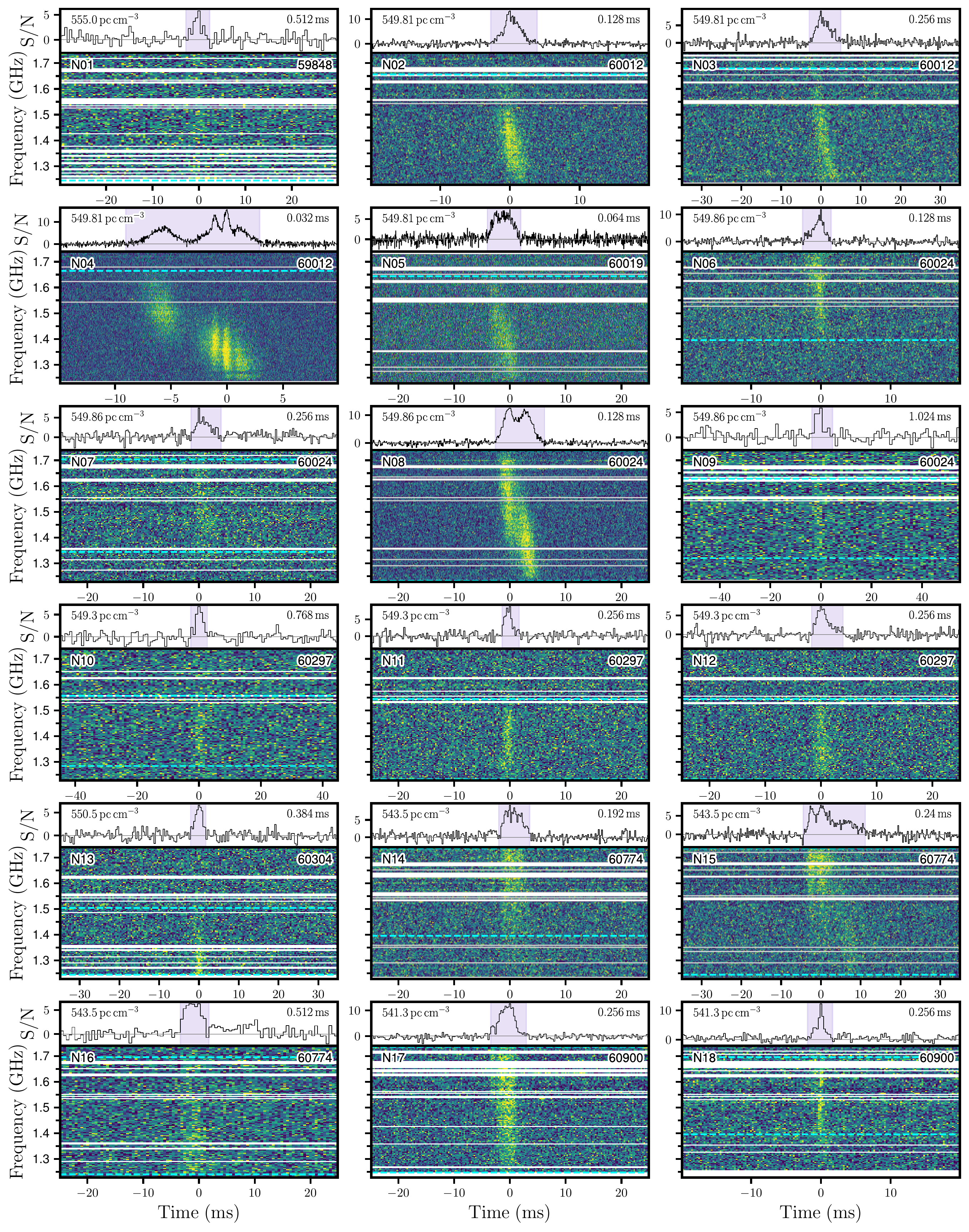}
\caption{\textbf{Dynamic spectra (bottom sub-panels) and frequency-integrated temporal profiles (top sub-panels) of the $18$ bursts} that were detected with the Nan\c{c}ay Radio Telescope. Intra-channel (coherent) dedispersion was applied during the observation using a DM that depends on the MJD of the observation (Table~\ref{tab:nancay_obs_log}). The inter-channel (incoherent) dedispersion values that were used for each burst are shown in the top left of every panel. All bursts are shown with a frequency resolution of $4$\,MHz and their time resolutions used for plotting are shown in the top right corners. The burst label and the MJD of the observations are shown in the top left and right corners of each dynamic spectrum, respectively. Horizontal white bands are frequency channels that are flagged because of RFI. Horizontal dashed cyan lines are the manually determined frequency ranges over which the dynamic spectra are averaged to create the timeseries. The manually determined vertical regions in the timeseries indicate the start and stop times of the bursts. For visual purposes the colour map limits have been set to the $7.5^{\mathrm{th}}$ and $99^{\mathrm{th}}$ percentile of each dynamic spectrum.}
\label{fig:family_plot_NRT}
\end{figure*}

We show the dynamic spectra and corresponding timeseries of the $18$ bursts detected by NRT in Figure~\ref{fig:family_plot_NRT}. For every burst we determine its time of arrival, temporal width, bandwidth, peak S/N, fluence, spectral luminosity, and peak flux density; and tabulate the results in Table~\ref{tab:burst_properties_NRT}.

Table~\ref{tab:nancay_obs_log} shows every observation of \rone with the Nan\c{c}ay Radio Telescope. Due to its design (Kraus-type transit telescope), a typical NRT observation of \rone lasts for $\sim$$50$\,minutes. Every observation has $512$\,MHz of bandwidth, with $4$\,MHz channels that have been coherently dedispersed to a particular value (Table~\ref{tab:nancay_obs_log}). Most observations are at L-band ($\sim$$1.4$\,GHz), with the occasional S-band observation ($\sim$$2.5$\,GHz).

\begin{table*}[ht]
    \centering
    \caption{Burst properties of the 18 bursts that were detected with the NRT.}
\begin{tabular}{c S[table-format=5.10,group-digits=none] S[table-format=3.2(3),separate-uncertainty=true] S[table-format=2.1,group-digits=none] S[table-format=2.3,group-digits=none] S[table-format=3.0,group-digits=none] S[table-format=1.2,group-digits=none] S[table-format=1.2,group-digits=none] S[table-format=2.2,group-digits=none]}
     \hline
     \hline
{Burst} & {ToA$^{\text{a}}$} & {DM$^{\text{b}}$} & {Peak$^{\text{c}}$} & {Width$^{\text{d}}$} & {BW$^{\text{e}}$} & {Fluence$^{\text{f}}$} & {Peak Flux$^{\text{f}}$} & {Spectral Luminosity$^{\text{f,g}}$} \\
{label} & {[MJD]} & {[pc\,cm$^{-3}$]} & {S/N} & {[ms]} & {[MHz]} & {[Jy\,ms]} & {Density [Jy]} & {[$10^{31}$ erg s$^{-1}$ Hz$^{-1}$]} \\ \midrule
N01 & 59848.2017213401 & 555 \pm 2       &  5.8 & 5.12 & 496 & 0.46 & 0.24 & 7.16 \\
N02 & 60012.7552257359 & 549.81          & 13.3 & 6.656 & 428 & 2.52 & 1.05 & 29.82 \\
N03 & 60012.7671579221 & 549.81          &  9.1 & 7.936 & 452 & 1.59 & 0.5 & 15.86 \\
N04 & 60012.7704943603 & 549.81 \pm 0.23 & 15.6 & 11.968 & 436 & 8.34 & 2.38 & 54.98 \\
N05 & 60019.7429292331 & 549.81          &  7.4 & 6.016 & 416 & 2.84 & 0.85 & 37.25 \\
N06 & 60024.7037589356 & 549.86          & 11.8 & 4.096 & 340 & 1.5 & 1.06 & 28.8 \\
N07 & 60024.7191847436 & 549.86          &  7.4 & 5.376 & 360 & 0.91 & 0.46 & 13.29 \\
N08 & 60024.7337595134 & 549.86 \pm 0.35 & 12.5 & 8.832 & 504 & 4.46 & 0.91 & 39.88 \\
N09 & 60024.7338868950 & 549.86          &  5.9 & 8.192 & 308 & 0.69 & 0.19 & 6.64 \\
N10 & 60297.9863805427 & 549.3           &  6.8 & 5.376 & 272 & 0.74 & 0.27 & 10.79 \\
N11 & 60297.9959326822 & 549.3           &  8.8 & 3.072 & 312 & 0.87 & 0.56 & 22.31 \\
N12 & 60298.0025116588 & 549.3 \pm 0.8   &  7.3 & 5.632 & 312 & 1.16 & 0.46 & 16.22 \\
N13 & 60304.9834779159 & 550.5 \pm 1.5   &  6.6 & 3.84 & 260 & 0.9 & 0.4 & 18.53 \\
N14 & 60774.6625003211 & 543.5 \pm 0.7   &  9.2 & 5.568 & 340 & 2.18 & 0.71 & 30.9 \\
N15 & 60774.6798939803 & 543.5           &  7.8 & 15.6 & 492 & 3.1 & 0.43 & 15.69 \\
N16 & 60774.6842242806 & 543.5           &  6.3 &  5.12 & 456 & 0.91 & 0.25 & 14.0 \\
N17 & 60900.3159690988 & 541.3 \pm 0.5   & 12.5 & 6.4 & 488 & 2.16 & 0.7 & 26.66 \\
N18 & 60900.3403471921 & 541.3           & 12.3 & 3.584 & 300 & 0.97 & 0.87 & 21.41 \\
\bottomrule
\multicolumn{9}{l}{$^{\text{a}}$ Corrected to the Solar System Barycentre to infinite frequency assuming the per-burst dispersion measure from the table,} \\
\multicolumn{9}{l}{\quad a reference frequency of $1738$\,MHz, a dispersion measure constant of $1/(2.41 \times10^{-4})$\,MHz$^{2}$\,pc$^{-1}$\,cm$^{3}$\,s,}  \\
\multicolumn{9}{l}{\quad a source of position of \evnburstslband\ and a} \\
\multicolumn{9}{l}{\quad NRT position of $X = 4324165.81$\,m, $Y = 165927.11$\,m and $Z = 4670132.83$\,m.} \\
\multicolumn{9}{l}{\quad The times quoted are dynamical times (TDB).} \\
\multicolumn{9}{l}{$^{\text{b}}$ We determine the DM for a subset of the NRT bursts, which are indicated with DM values that have a corresponding error.} \\
\multicolumn{9}{l}{\quad Bursts for which we do not determine a DM we use the determined DM that is closest in time. For all bursts the quoted} \\
\multicolumn{9}{l}{\quad DM is the DM that is used in the incoherent dedispersion in Figure~\ref{fig:family_plot_NRT}.} \\
\multicolumn{9}{l}{$^{\text{c}}$ The peak value of the timeseries, as shown in Figure~\ref{fig:family_plot_NRT}.} \\
\multicolumn{9}{l}{$^{\text{d}}$ Manually determined time span of the burst, shown as the highlighted region in the timeseries in Figure~\ref{fig:family_plot_NRT}.} \\
\multicolumn{9}{l}{$^{\text{e}}$ Manually determined frequency span of the burst, i.e., the difference between the dashed cyan lines in Figure~\ref{fig:family_plot_NRT}.} \\
\multicolumn{9}{l}{$^{\text{f}}$ We estimate a (conservative) error of 20\% for these values, which is dominated by the uncertainty in the system} \\
\multicolumn{9}{l}{\quad equivalent flux density (SEFD) of the Nan\c{c}ay Radio Telescope ($\mathrm{T}_{\mathrm{sys}} = 35$\,K and $\mathrm{Gain} = 1.4$\,K/Jy).} \\
\multicolumn{9}{l}{$^{\text{g}}$ Using Equation~5 from \citet{ouldboukattine_2024_arxiv} and assuming a luminosity distance of $972$\,Mpc ($z = 0.193$).} \\
    \end{tabular}
    \label{tab:burst_properties_NRT}
\end{table*}

\begin{table*}[ht]
\small
    \centering
    \caption{Nan\c{c}ay Radio Telescope observations.}\label{tab:nancay_obs_log}
    \begin{tabular}{clcccc||clcccc}
    \hline \hline
    MJD & Observation & Freq.$^{\mathrm{a}}$ & Duration & DM$^{\mathrm{b}}$ & Num. of &  MJD & Observation & Freq. & Duration & DM              & Num. of  \\
        & date        & [MHz]                & [min.]   & [pc\,cm$^{-3}$]   & bursts  &      & date        & [MHz] & [min.]   & [pc\,cm$^{-3}$] & bursts   \\
    \hline
59616 & 2022 Feb. 06 & 1484 & 46.9 & 564.0 & 0 & 60350 & 2024 Feb. 10 & 1484 & 51.2 & 564.0   & 0 \\
59679 & 2022 Apr. 10 & 1484 & 49.6 & 564.0 & 0 & 60358 & 2024 Feb. 18 & 1484 & 51.2 & 564.0   & 0 \\
59683 & 2022 Apr. 14 & 1484 & 55.1 & 564.0 & 0 & 60368 & 2024 Feb. 28 & 1484 & 51.2 & 564.0   & 0 \\
59688 & 2022 Apr. 19 & 1484 & 51.5 & 564.0 & 0 & 60378 & 2024 Mar. 09 & 1484 & 51.2 & 564.0   & 0 \\
59693 & 2022 Apr. 24 & 1484 & 51.0 & 564.0 & 0 & 60385 & 2024 Mar. 16 & 1484 & 53.4 & 564.0   & 0 \\
59697 & 2022 Apr. 28 & 1484 & 51.5 & 564.0 & 0 & 60392 & 2024 Mar. 23 & 1484 & 51.2 & 564.0   & 0 \\
59703 & 2022 May. 04 & 1484 & 51.5 & 564.0 & 0 & 60399 & 2024 Mar. 30 & 1484 & 51.2 & 564.0   & 0 \\
59706 & 2022 May. 07 & 1484 & 55.2 & 564.0 & 0 & 60406 & 2024 Apr. 06 & 1484 & 51.2 & 564.0   & 0 \\
59708 & 2022 May. 09 & 1484 & 51.5 & 564.0 & 0 & 60413 & 2024 Apr. 13 & 1484 & 51.2 & 564.0   & 0 \\
59716 & 2022 May. 17 & 1484 & 33.4 & 564.0 & 0 & 60420 & 2024 Apr. 20 & 1484 & 51.2 & 564.0   & 0 \\
59731 & 2022 Jun. 01 & 1484 & 51.4 & 564.0 & 0 & 60427 & 2024 Apr. 27 & 1484 & 51.2 & 564.0   & 0 \\
59738 & 2022 Jun. 08 & 1484 & 60.3 & 564.0 & 0 & 60436 & 2024 May. 06 & 1484 & 62.8 & 564.0   & 0 \\
59745 & 2022 Jun. 15 & 1484 & 53.7 & 564.0 & 0 & 60443 & 2024 May. 13 & 1484 & 53.4 & 564.0   & 0 \\
59757 & 2022 Jun. 27 & 1484 & 50.5 & 564.0 & 0 & 60450 & 2024 May. 20 & 1484 & 62.8 & 564.0   & 0 \\
59763 & 2022 Jul. 03 & 1484 & 62.8 & 564.0 & 0 & 60474 & 2024 Jun. 13 & 1484 & 56.6 & 564.0   & 0 \\
59848 & 2022 Sep. 26 & 1484 & 51.4 & 564.0 & 1 & 60481 & 2024 Jun. 20 & 1484 & 62.8 & 564.0   & 0 \\
59855 & 2022 Oct. 03 & 1484 & 62.9 & 564.0 & 0 & 60488 & 2024 Jun. 27 & 1484 & 62.8 & 564.0   & 0 \\
59862 & 2022 Oct. 10 & 1484 & 51.4 & 564.0 & 0 & 60495 & 2024 Jul. 04 & 1484 & 51.2 & 564.0   & 0 \\
59864 & 2022 Oct. 12 & 1484 & 51.4 & 564.0 & 0 & 60502 & 2024 Jul. 11 & 1484 & 51.2 & 564.0   & 0 \\
59874 & 2022 Oct. 22 & 1484 & 60.1 & 564.0 & 0 & 60509 & 2024 Jul. 18 & 1484 & 44.4 & 564.0   & 0 \\
59876 & 2022 Oct. 24 & 1484 & 51.4 & 564.0 & 0 & 60527 & 2024 Aug. 05 & 1484 & 51.2 & 564.0   & 0 \\
60005 & 2023 Mar. 02 & 1484 & 51.4 & 564.0 & 0 & 60534 & 2024 Aug. 12 & 1484 & 62.8 & 564.0   & 0 \\
60012 & 2023 Mar. 09 & 1484 & 51.4 & 564.0 & 3 & 60541 & 2024 Aug. 19 & 1484 & 33.5 & 564.0   & 0 \\
60019 & 2023 Mar. 16 & 1484 & 55.1 & 564.0 & 1 & 60548 & 2024 Aug. 26 & 1484 & 51.2 & 564.0   & 0 \\
60022 & 2023 Mar. 19 & 1484 & 42.4 & 564.0 & 0 & 60596 & 2024 Oct. 13 & 1484 & 62.6 & 564.0   & 0 \\
60024 & 2023 Mar. 21 & 1484 & 63.0 & 564.0 & 4 & 60604 & 2024 Oct. 21 & 1484 & 44.4 & 564.0   & 0 \\
60027 & 2023 Mar. 24 & 1484 & 38.3 & 564.0 & 0 & 60611 & 2024 Oct. 28 & 1484 & 51.2 & 564.0   & 0 \\
60028 & 2023 Mar. 25 & 1484 & 38.3 & 564.0 & 0 & 60618 & 2024 Nov. 04 & 1484 & 44.4 & 564.0   & 0 \\
60029 & 2023 Mar. 26 & 1484 & 63.0 & 564.0 & 0 & 60625 & 2024 Nov. 11 & 1484 & 44.4 & 564.0   & 0 \\
60031 & 2023 Mar. 28 & 1484 & 63.0 & 564.0 & 0 & 60632 & 2024 Nov. 18 & 1484 & 51.2 & 564.0   & 0 \\
60032 & 2023 Mar. 29 & 1484 & 58.5 & 564.0 & 0 & 60646 & 2024 Dec. 02 & 2539 & 51.3 & 564.0   & 0 \\
60033 & 2023 Mar. 30 & 1484 & 51.4 & 564.0 & 0 & 60652 & 2024 Dec. 08 & 2509 & 45.8 & 564.0   & 0 \\
60035 & 2023 Apr. 01 & 1484 & 63.0 & 564.0 & 0 & 60661 & 2024 Dec. 17 & 2539 & 53.2 & 564.0   & 0 \\
60037 & 2023 Apr. 03 & 1484 & 62.3 & 564.0 & 0 & 60668 & 2024 Dec. 24 & 1484 & 44.4 & 564.0   & 0 \\
60039 & 2023 Apr. 05 & 1484 & 55.1 & 564.0 & 0 & 60675 & 2024 Dec. 31 & 1484 & 44.4 & 564.0   & 0 \\
60046 & 2023 Apr. 12 & 1484 & 53.6 & 564.0 & 0 & 60682 & 2025 Jan. 07 & 2509 & 46.3 & 564.0   & 0 \\
60049 & 2023 Apr. 15 & 1484 & 65.0 & 564.0 & 0 & 60689 & 2025 Jan. 14 & 2509 & 46.3 & 564.0   & 0 \\
60052 & 2023 Apr. 18 & 1484 & 54.7 & 564.0 & 0 & 60696 & 2025 Jan. 21 & 1484 & 44.4 & 564.0   & 0 \\
60057 & 2023 Apr. 23 & 1484 & 40.1 & 564.0 & 0 & 60703 & 2025 Jan. 28 & 1484 & 44.4 & 564.0   & 0 \\
60067 & 2023 May. 03 & 1484 & 55.1 & 564.0 & 0 & 60710 & 2025 Feb. 04 & 1484 & 51.2 & 555.0   & 0 \\
60075 & 2023 May. 11 & 1484 & 63.0 & 564.0 & 0 & 60717 & 2025 Feb. 11 & 1484 & 44.4 & 555.0   & 0 \\
60078 & 2023 May. 14 & 1484 & 46.9 & 564.0 & 0 & 60724 & 2025 Feb. 18 & 1484 & 44.4 & 555.0   & 0 \\
60082 & 2023 May. 18 & 1484 & 51.4 & 564.0 & 0 & 60731 & 2025 Feb. 25 & 1484 & 44.4 & 555.0   & 0 \\
60089 & 2023 May. 25 & 1484 & 57.3 & 564.0 & 0 & 60738 & 2025 Mar. 04 & 1484 & 44.4 & 555.0   & 0 \\
60096 & 2023 Jun. 01 & 1484 & 51.4 & 564.0 & 0 & 60745 & 2025 Mar. 11 & 1484 & 44.4 & 555.0   & 0 \\
60104 & 2023 Jun. 09 & 1484 & 51.4 & 564.0 & 0 & 60752 & 2025 Mar. 18 & 1484 & 48.9 & 555.0   & 0 \\
60111 & 2023 Jun. 16 & 1484 & 51.4 & 564.0 & 0 & 60759 & 2025 Mar. 25 & 1484 & 51.2 & 555.0   & 0 \\
60221 & 2023 Oct. 04 & 1484 & 51.4 & 564.0 & 0 & 60767 & 2025 Apr. 02 & 1484 & 44.4 & 555.0   & 0 \\
60228 & 2023 Oct. 11 & 1484 & 51.4 & 564.0 & 0 & 60774 & 2025 Apr. 09 & 1484 & 44.4 & 555.0   & 3 \\
60236 & 2023 Oct. 19 & 1484 & 51.4 & 564.0 & 0 & 60789 & 2025 Apr. 24 & 1484 & 46.7 & 543.5   & 0 \\
60243 & 2023 Oct. 26 & 1484 & 51.4 & 564.0 & 0 & 60797 & 2025 May. 02 & 1484 & 26.4 & 543.5   & 0 \\
60250 & 2023 Nov. 02 & 1484 & 51.4 & 564.0 & 0 & 60804 & 2025 May. 09 & 1484 & 42.2 & 543.5   & 0 \\
60257 & 2023 Nov. 09 & 1484 & 51.4 & 564.0 & 0 & 60811 & 2025 May. 16 & 1484 & 39.9 & 545.0   & 0 \\
60264 & 2023 Nov. 16 & 1484 & 53.6 & 564.0 & 0 & 60818 & 2025 May. 23 & 1484 & 44.4 & 545.0   & 0 \\
60269 & 2023 Nov. 21 & 1484 & 63.0 & 564.0 & 0 & 60825 & 2025 May. 30 & 1484 & 44.4 & 545.0   & 0 \\
60275 & 2023 Nov. 27 & 1484 & 51.4 & 564.0 & 0 & 60832 & 2025 Jun. 06 & 1484 & 51.2 & 545.0   & 0 \\
60284 & 2023 Dec. 06 & 1484 & 51.4 & 564.0 & 0 & 60848 & 2025 Jun. 22 & 1484 & 60.0 & 545.0   & 0 \\
60290 & 2023 Dec. 12 & 1484 & 51.4 & 564.0 & 0 & 60858 & 2025 Jul. 02 & 1484 & 51.2 & 545.0   & 0 \\
60297 & 2023 Dec. 19 & 1484 & 51.4 & 564.0 & 3 & 60870 & 2025 Jul. 14 & 1484 & 51.2 & 545.0   & 0 \\
60304 & 2023 Dec. 26 & 1484 & 51.3 & 564.0 & 1 & 60884 & 2025 Jul. 28 & 1484 & 51.2 & 545.0   & 0 \\
60318 & 2024 Jan. 09 & 1484 & 33.5 & 564.0 & 0 & 60893 & 2025 Aug. 06 & 1484 & 51.2 & 545.0   & 0 \\
60325 & 2024 Jan. 16 & 1484 & 53.4 & 564.0 & 0 & 60900 & 2025 Aug. 13 & 1484 & 44.5 & 545.0   & 2 \\
60333 & 2024 Jan. 24 & 1484 & 51.2 & 564.0 & 0 & \multicolumn{6}{l}{\textbf{Totals: 125 observations, 106 hours, 18 bursts}} \\ \hline
        \multicolumn{10}{l}{$^{\mathrm{a}}$Every observation uses $512$\,MHz of bandwidth centred around this frequency, with $4$-MHz channels.} \\
        \multicolumn{10}{l}{$^{\mathrm{b}}$The dispersion measure value that was used for the coherent dedispersion during the observation.}
    \end{tabular}
\end{table*}

\section{Dispersion measure determination}
\label{app:dm_opt}

\subsection{EVN data}

\begin{figure}[ht]
\centering
\includegraphics[width=1.0\linewidth]{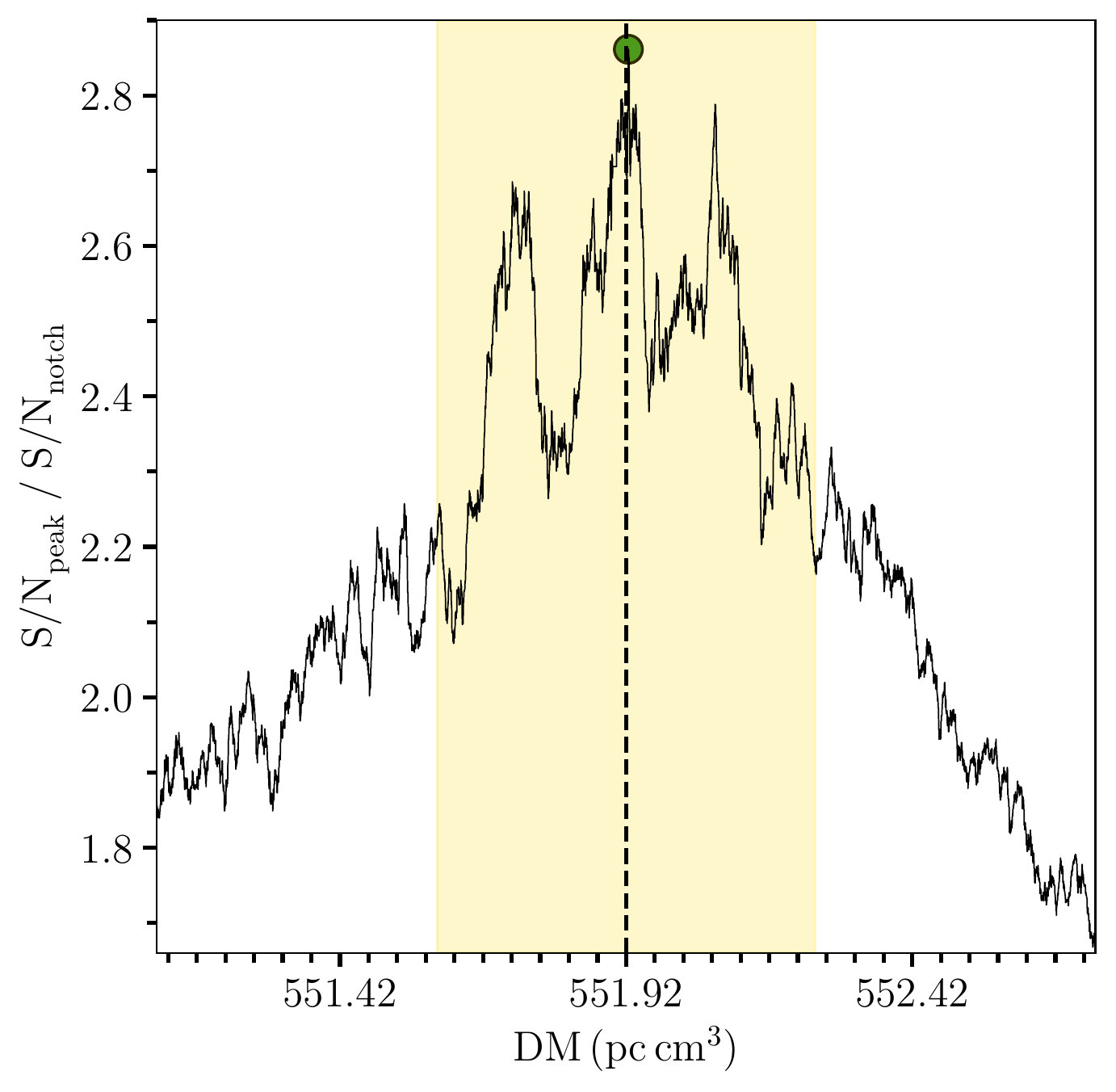}
\caption{\textbf{Dispersion measure (DM) determination}. The figure shows the ratio between the peak S/N of burst B$3$ and the S/N of the `notch' between the first and second burst component (Figure~\ref{fig:family_plot}, panel B$3$) as a function of DM. The `best' DM is where this ratio is maximized, $551.92$\,pc\,cm$^{-3}$, and the error is estimated to be $0.33$\,pc\,cm$^{-3}$ (highlighted region).}
\label{app-fig:dm_opt}
\end{figure}

We only use burst B$3$ (Figure~\ref{fig:family_plot}) from the Effelsberg radio telescope data to determine the DM of the EVN bursts, because burst B$3$ has both the highest fluence and lowest burst width (Table~\ref{tab:burst_properties}). We first use the search data to, by eye, determine roughly the `right' DM, $551.9$\,pc\,cm$^{-3}$. Next, we use \texttt{SFXC} to coherently dedisperse the data, within the channels, to that value and write out the Stokes~I (intensity) data as filterbank files with a time and frequency resolution of $4$\,$\upmu$s and $500$\,kHz, respectively. Next, we incoherently dedisperse the data for a range of DM values, from $550.9$\,pc\,cm$^{-3}$ to $552.9$\,pc\,cm$^{-3}$ in steps of $0.0005$\,pc\,cm$^{-3}$. The high time resolution of the data was chosen because the incoherent dedispersion can only shift the channels by an integer number of bins. For every iteration we mask RFI-affected channels, downsample the data to an effective time resolution of $80$\,$\upmu$s, and average over the frequency extent of the burst. This creates a timeseries in units of S/N, and for every timeseries we determine the ratio between the S/N of the peak and the S/N of the notch between the two burst components (Figure~\ref{fig:family_plot}, panel B$3$). The results are shown in Figure~\ref{app-fig:dm_opt}. The `best' DM is where this ratio is maximized, which occurs at $\mathrm{DM} = 551.92$\,pc\,cm$^{-3}$. Given the scatter in the ratio between S/Ns data points, we conservatively estimate the error on the DM to be $0.33$\,pc\,cm$^{-3}$ (the highlighted region in Figure~\ref{app-fig:dm_opt}).

\subsection{NRT data}
We determine the DM for a subset of the bursts that were detected by NRT using a similar strategy as previously mentioned, with the difference that NRT does not record baseband data. Therefore, we can only apply integer bin shifts to the $16$\,$\upmu$s-resolution data during the dedispersion before downsampling in time and averaging in frequency. Bursts for which we do not determine a DM, are incoherently dedispersed to the determined DM value that is closest in time. The results are shown in Figure~\ref{fig:dm_evolution} and tabulated in Table~\ref{tab:burst_properties_NRT}.

As previously mentioned, NRT does not record baseband data but rather performs coherent dedispersion on the streaming voltage data during the observation. A mismatch between the true DM value and the DM that was used during the observation will lead to intra-channel temporal smearing. Given the $4$-MHz wide channels of the NRT data, this smearing is $6$\,$\upmu$s and $18$\,$\upmu$s per DM unit at the top and bottom of the NRT L-band, respectively. There is, at most, a $\sim$$14$\,pc\,cm$^{-3}$ difference between the true DM and the DM value that was used for coherent dedispersion (e.g., burst N$04$). This leads to a temporal smearing of $\sim$$250$\,$\upmu$s at the bottom of the band. Given that all the bursts, including their individual components, are wider than this, we expect that this temporal smearing will not substantially affect the DM determination.

\section{Burst and PRS localisation images}
\label{app:burst_loc_images}

\begin{figure*}[ht]
\centering
\includegraphics[width=1.0\linewidth]{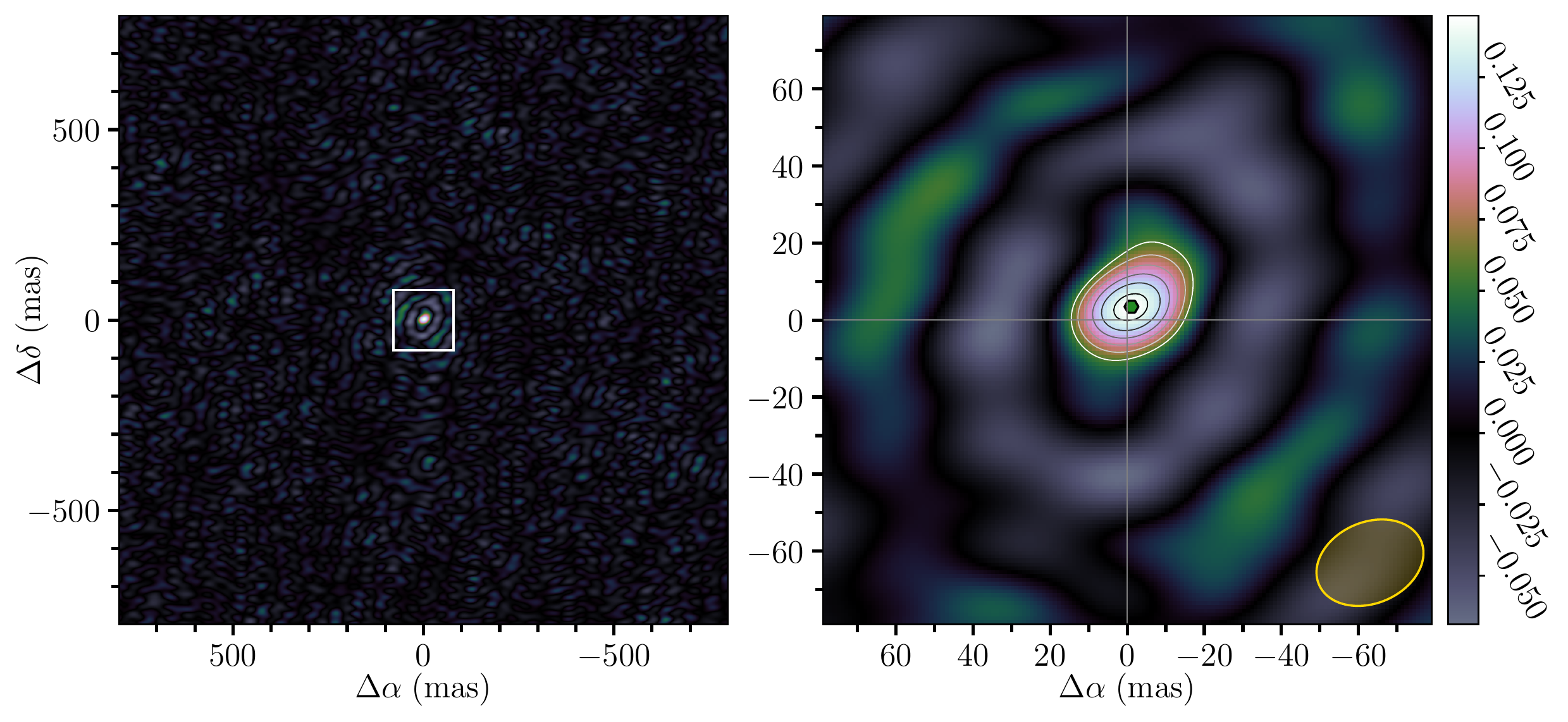}
\caption{Zoomed-out version of the right panel of Figure~\ref{fig:bursts_prs}, without limits on the colour map. The colour bar is in units of mJy/beam.}
\label{app-fig:prs_zoom_out}
\end{figure*}

\begin{figure*}[ht]
\centering
\includegraphics[width=1.0\linewidth]{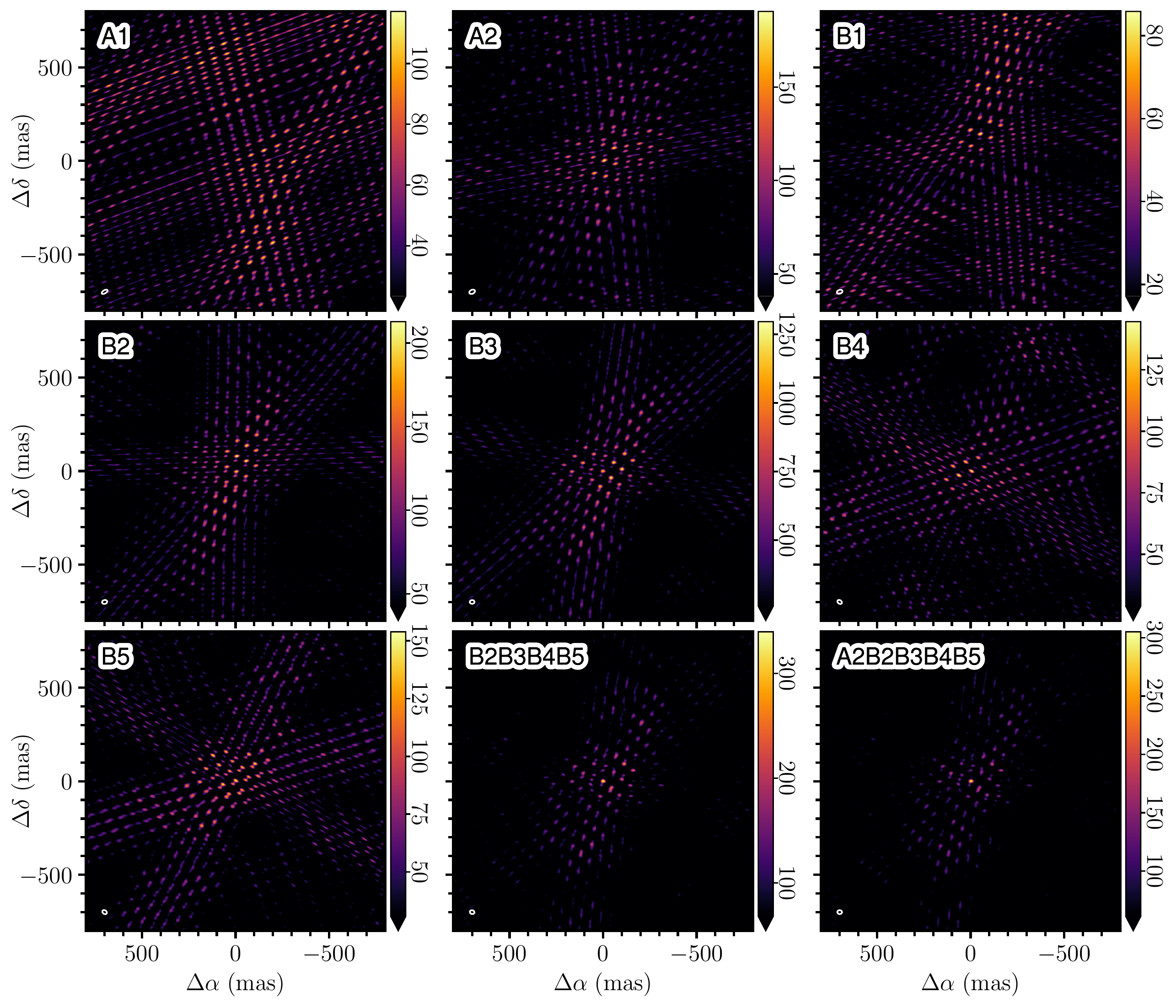}
\caption{\textbf{Zoomed-out version of Figure~\ref{fig:bursts_dirty_zoom}}, showing $800\times800$\,mas panels. See Figure~\ref{app-fig:burst_dirty_no_cap} for a version of this image without limits on the colour maps. The colour bars are in units of mJy/beam and the limits are $20$--$100$\,\% of the maximum value.}
\label{app-fig:burst_dirty_all}
\end{figure*}

\begin{figure*}[ht]
\centering
\includegraphics[width=1.0\linewidth]{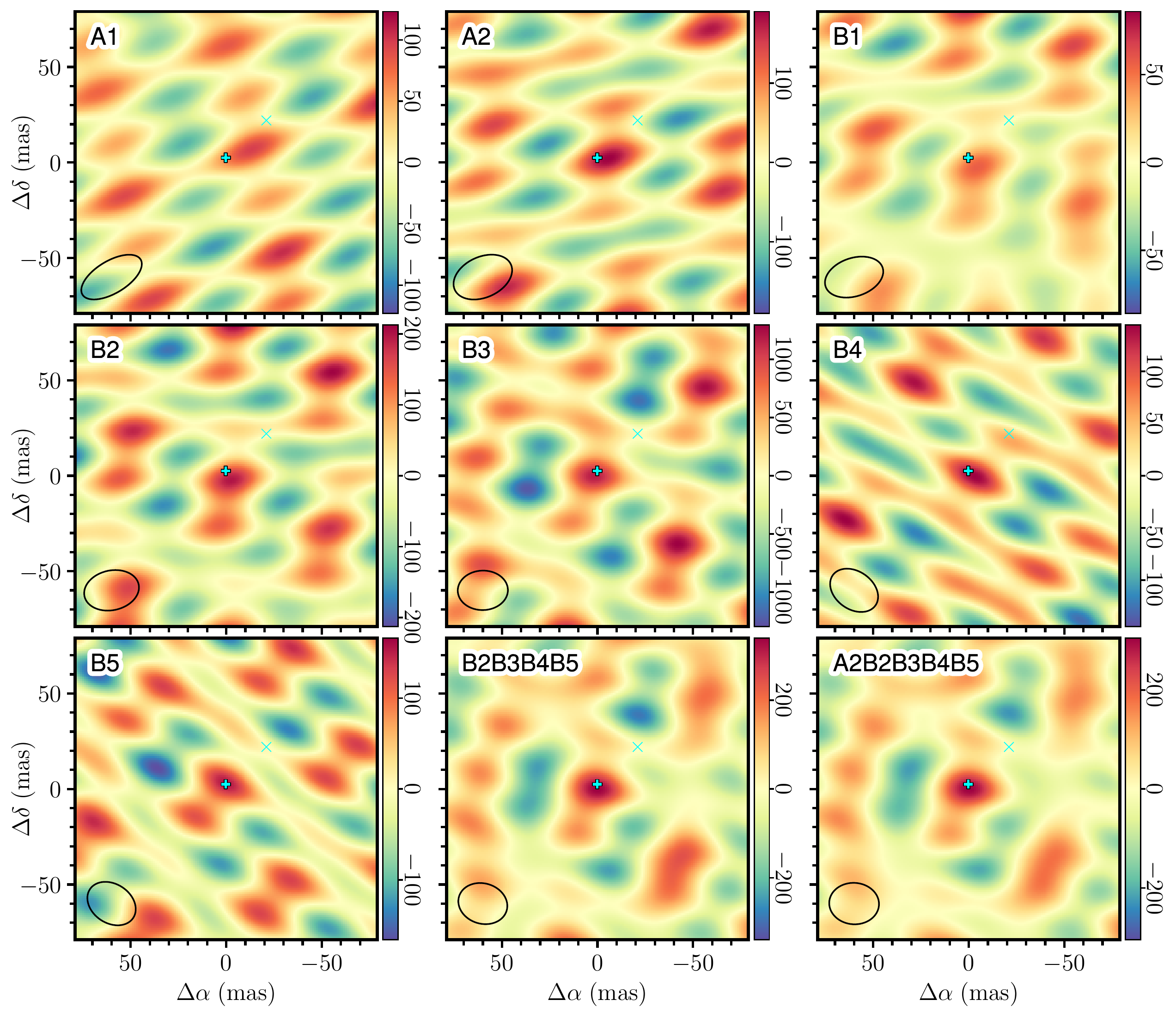}
\caption{Same as Figure~\ref{fig:bursts_dirty_zoom}, but without limits on the colour map. The colour bars are in units of mJy/beam.}
\label{app-fig:burst_dirty_no_cap_zoom}
\end{figure*}

\begin{figure*}[ht]
\centering
\includegraphics[width=1.0\linewidth]{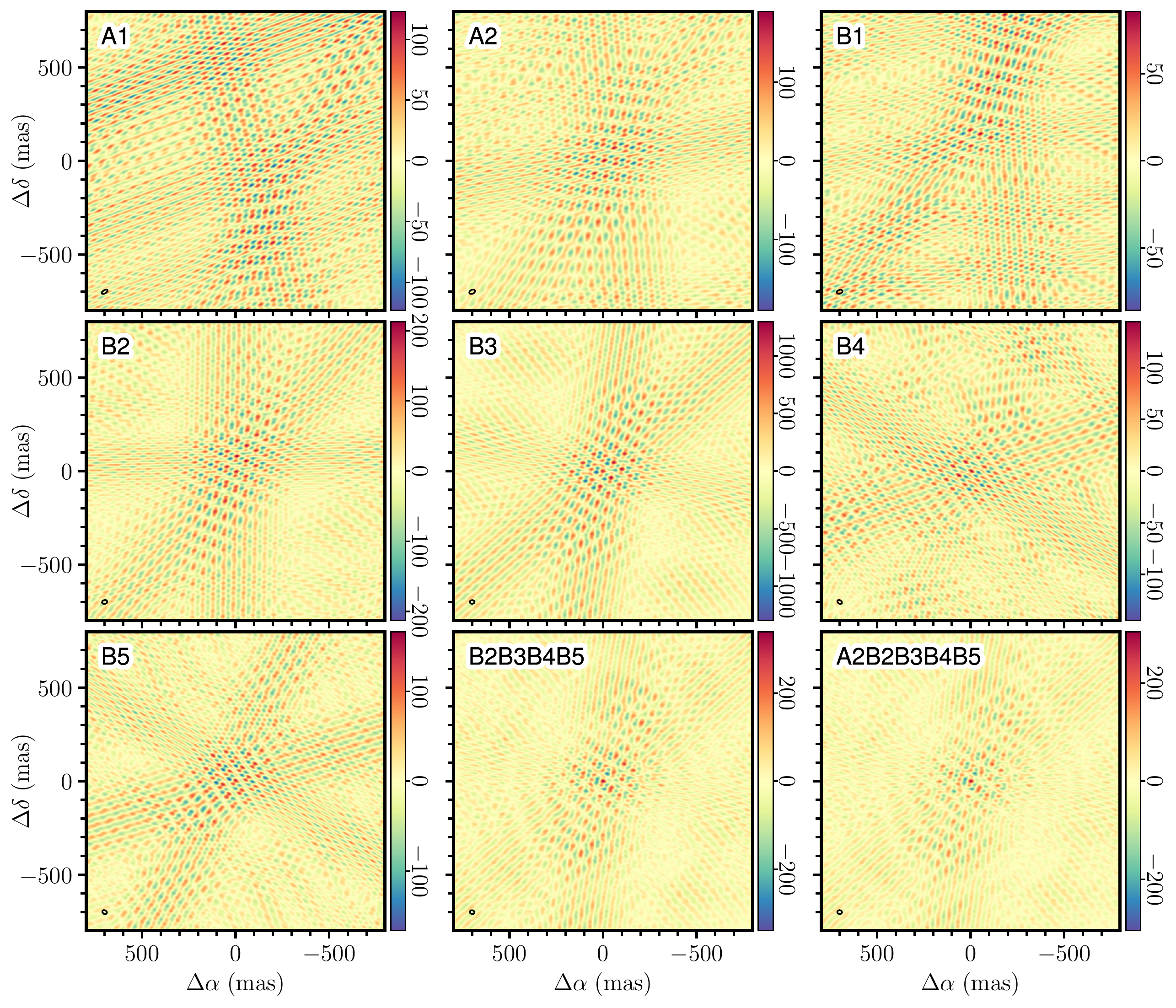}
\caption{Same as Figure~\ref{app-fig:burst_dirty_all}, but without limits on the colour map. The colour bars are in units of mJy/beam.}
\label{app-fig:burst_dirty_no_cap}
\end{figure*}

\begin{figure*}[ht]
\centering
\includegraphics[width=1.0\linewidth]{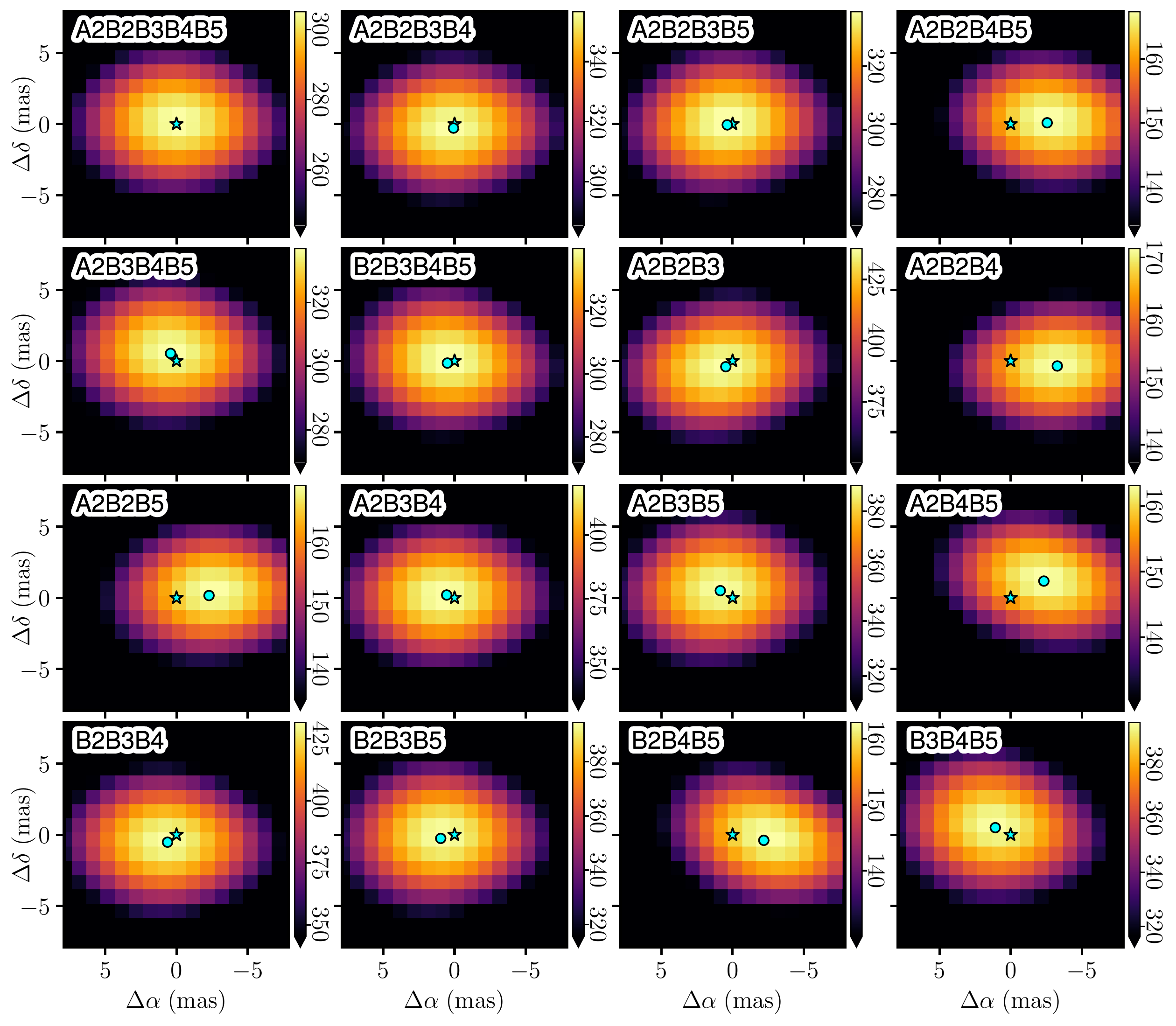}
\caption{\textbf{Demonstration of small, mas-level, shifts in the burst positions}, depending on which bursts are used in the imaging. The best position, which is derived from the combination of A$2$, B$2$, B$3$, B$4$, and B$5$, is shown as a cyan star in all panels. Every combination with $3$ and $4$ bursts is imaged and their respective dirty maps are fitted to get their best positions, which are indicated as cyan circles. These positions are shown in Figure~\ref{fig:position_scatter}. Note that the extent of the panels is much smaller than the size of the synthesised beams ($\sim$$25\times25$\,mas). The colour bars are in units of mJy/beam and the limits are $80$--$100$\,\% of the maximum value of every panel.}
\label{app-fig:burst_combinations}
\end{figure*}

\begin{figure*}[ht]
\centering
\includegraphics[width=1.0\linewidth]{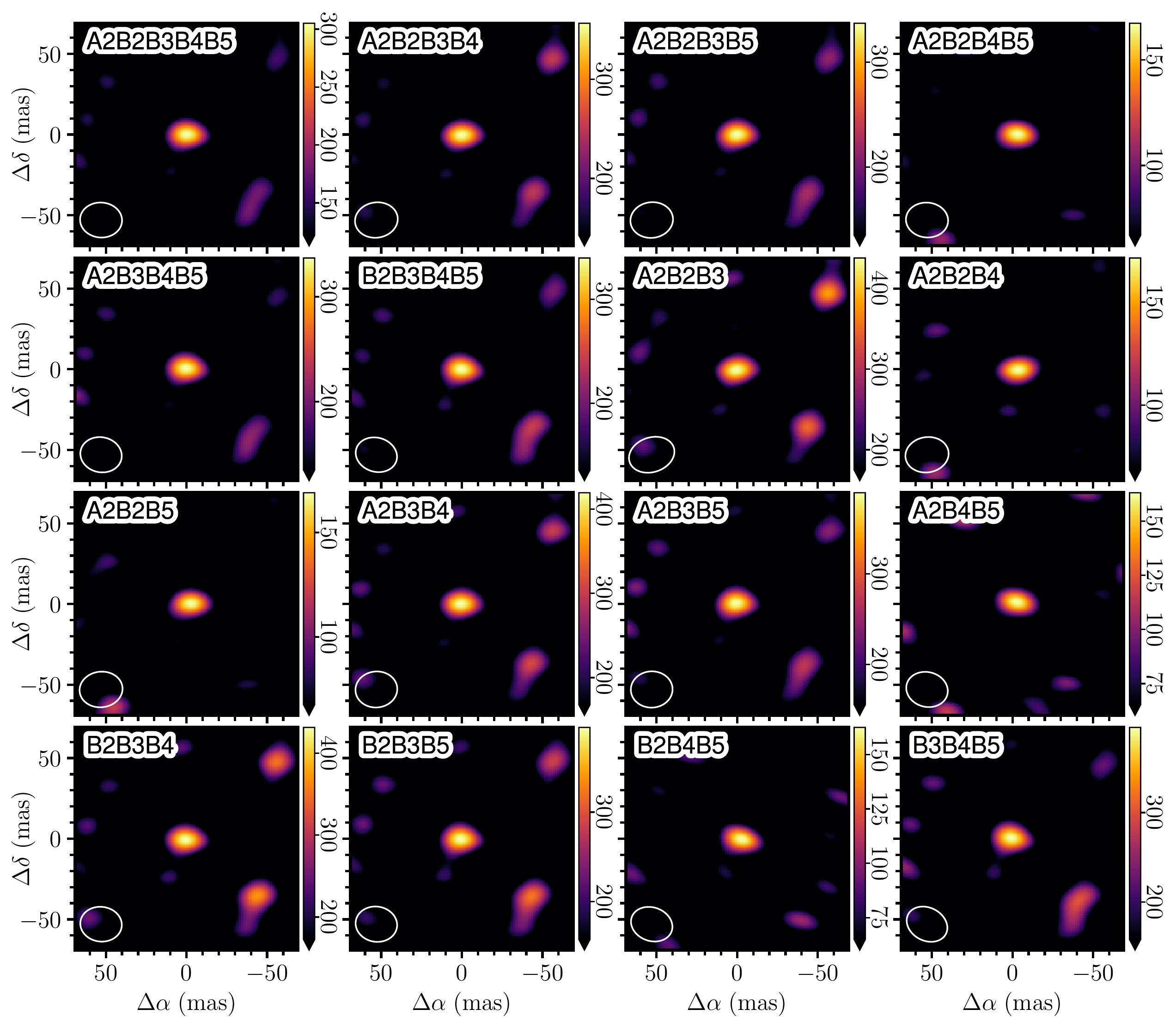}
\caption{\textbf{Solving `sidelobe ambiguity'}. We explore every combination that uses $5$ bursts (top left), $4$ bursts, and $3$ bursts (omitting bursts A$1$ and B$1$ due to their low S/N). It is clear that, in this case, every combination that uses 3 or more bursts results in an image without any ambiguity on the burst positions due to sidelobe structure. The colour bars are in units of mJy/beam and the white ellipses in the bottom left corners show the FWHM of the synthesised beams.}
\label{app-fig:no_sidelobe_ambiguity}
\end{figure*}

Here we provide Figures~\ref{app-fig:prs_zoom_out}, \ref{app-fig:burst_dirty_all}, \ref{app-fig:burst_dirty_no_cap_zoom}, \ref{app-fig:burst_dirty_no_cap}, \ref{app-fig:burst_combinations}, \ref{app-fig:no_sidelobe_ambiguity} that further illustrate the robustness of the astrometric results. Figure~\ref{app-fig:prs_zoom_out} shows the $800\times800$\,mas field around the PRS detection without any limits on the colour map to show that it is the only clearly detected radio source in this field. Figures~\ref{app-fig:burst_dirty_all}, \ref{app-fig:burst_dirty_no_cap_zoom} and \ref{app-fig:burst_dirty_no_cap} show different variations of Figure~\ref{fig:bursts_dirty_zoom} (zoom in/out, with and without limits on the colour map) to show the cross fringe pattern. The purpose of Figure~\ref{app-fig:burst_combinations} is to show that the position of the bursts can shift by $\sim$$2$\,mas, depending on which bursts are used in the imaging process. Finally, Figure~\ref{app-fig:no_sidelobe_ambiguity} shows that, in this case, sidelobe ambiguity is solved when $3$ or more bursts are used.

\section{NRT Polarimetry}
\label{app-sec:pol}

\begin{figure}[ht]
\centering
\includegraphics[width=1.0\linewidth]{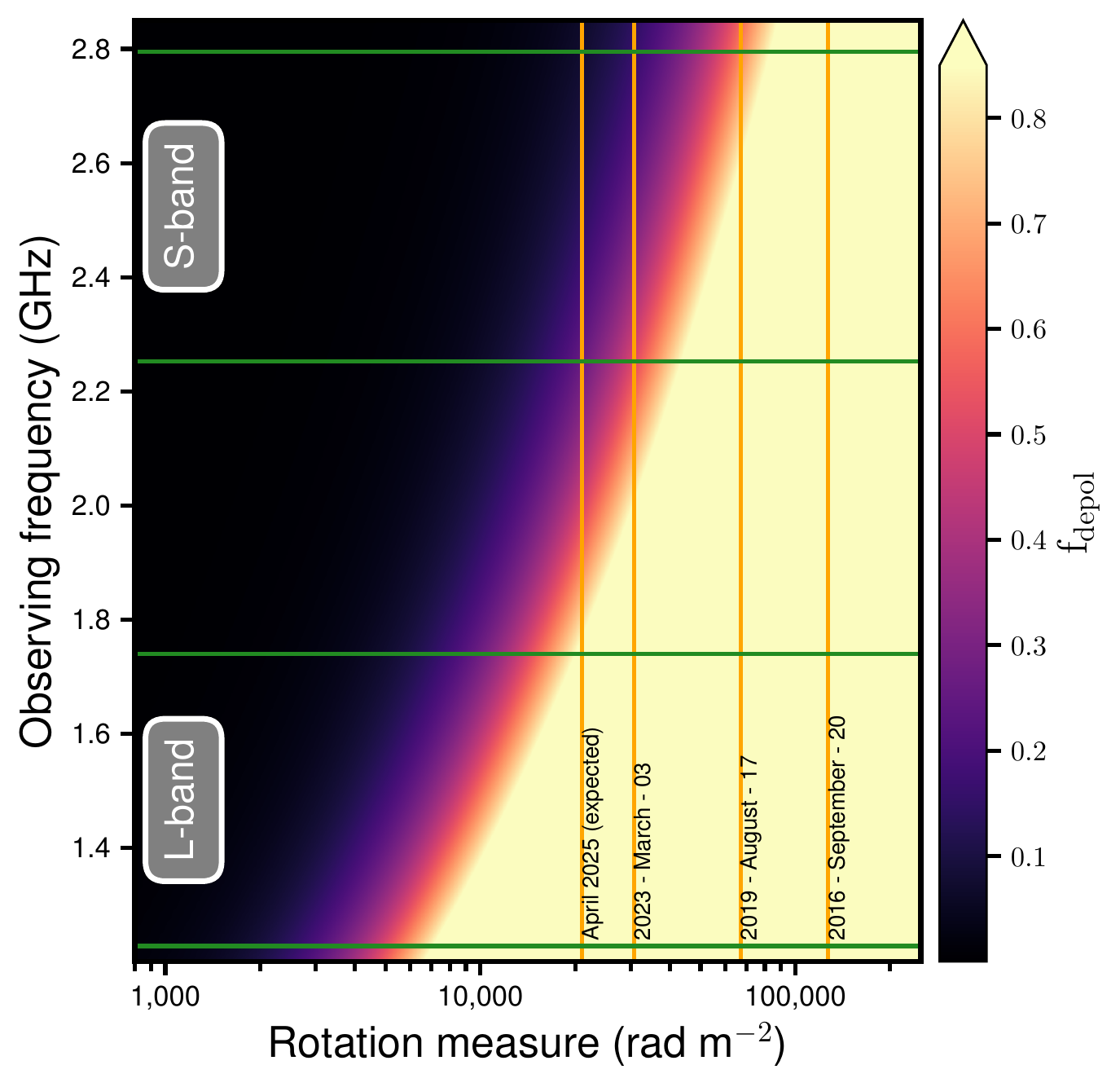}
\caption{\textbf{The intra-channel depolarisation fraction} as a function of absolute Faraday RM and observing frequency for the Nan\c{c}ay Radio Telescope, assuming a channel width of $4$\,MHz. The regions between the solid horizontal green lines indicate the frequency ranges of the L-band and S-band observations. Vertical orange lines are placed at RMs of $\sim$$127 \times 10^{3}$\,rad\,m$^{-2}$ \citep[the earliest RM measurement of \rone;][]{Plavin_2022_MNRAS}, $\sim$$67 \times 10^{3}$\,rad\,m$^{-2}$ \citep[the last RM measurement of \rone with the Arecibo telescope;][]{hilmarsson_2021_apjl}, $\sim$$31 \times 10^{3}$\,rad\,m$^{-2}$ \citep[the most recent RM measurement of \rone;][]{wang_2025_arxiv} and at $\sim$$21 \times 10^{3}$\,rad\,m$^{-2}$, which is the expected RM of the source (in April 2025) based on the $\Delta$RM of \citet{wang_2025_arxiv}.}
\label{app-fig:depol}
\end{figure}

A linearly polarised signal will rotate if it passes through a magnetised plasma. The intra-channel Faraday rotation is given by \citep{Michilli_2018_Nature}
\begin{equation}
    \Delta \theta = \mathrm{RM}_{\mathrm{obs}} c^{2} \nu^{-3}_{c} \Delta \nu,
\end{equation}
where $c$ is the speed of light, $\nu_{c}$ is the observing frequency, $\Delta \nu$ is the channel width and $\mathrm{RM}_{\mathrm{obs}}$ is the Faraday RM in the observer frame. To compensate for the high time resolution ($16$\,$\upmu$s) of the NRT data, the data is channelized to relatively wide channels, each $4$\,MHz wide. These large channel widths lead to significant depolarisation within a frequency channel, if the (absolute) RM of an astrophysical source is large. The depolarisation fraction is given by \citep{Michilli_2018_Nature}
\begin{equation}
    f_{\mathrm{depol}} = 1 - \left[ \sin \left( 2 \Delta \theta \right) / \ 2 \Delta \theta \right].
\end{equation}
The depolarisation fraction is illustrated in Figure~\ref{app-fig:depol} as a function of frequency and |RM|, assuming a channel width of $4$\,MHz. It can be seen that, for NRT L-band observations, depolarisation due to finite channel width becomes a problem if $\left| \mathrm{RM} \right| \gtrsim 3,000$\,rad\,m$^{-2}$. \rone has an extremely high RM, which is known to decrease from $\sim$$127 \times 10^{3}$ \,rad\,m$^{-2}$ to $\sim$$31 \times 10^{3}$ \,rad\,m$^{-2}$ \citep[quoted RM values are in the observer frame;][]{wang_2025_arxiv,Michilli_2018_Nature,Plavin_2022_MNRAS,hilmarsson_2021_apjl}. Based on the $\Delta$RM from \citet{wang_2025_arxiv} we expect that \rone had an RM of $\sim$$21 \times 10^{3}$\,rad\,m$^{-2}$ in April 2025. Therefore we do not perform any polarimetric investigations with the NRT data for the L-band bursts. No bursts at S-band were found during the NRT observations.

\end{document}